\documentclass[12pt]{article}

\usepackage{axodraw}

\makeatletter

\def\paragraph{\@startsection{paragraph}{4}{\z@}{+2.00ex plus
 +1ex minus +.2ex}{1.5ex plus .2ex}{\it\normalsize}}

\def\section{\@startsection {section}{1}{\z@}{+3.0ex plus +1ex minus
  +.2ex}{2.3ex plus .2ex}{\normalsize\bf\boldmath}}
\def\subsection{\@startsection{subsection}{2}{\z@}{+2.5ex plus +1ex
minus +.2ex}{1.5ex plus .2ex}{\normalsize\bf\boldmath}}
\def\subsubsection{\@startsection{subsubsection}{3}{\z@}{+3.25ex plus
 +1ex minus +.2ex}{1.5ex plus .2ex}{\normalsize\it}}

\expandafter\ifx\csname mathrm\endcsname\relax\def\mathrm#1{{\rm #1}}\fi

\renewcommand{\theequation}{\thesection.\arabic{equation}}
\newcounter{saveeqn}

\@addtoreset{equation}{section}

\newcount\@tempcntc
\def\@citex[#1]#2{\if@filesw\immediate\write\@auxout{\string\citation{#2}}\fi
  \@tempcnta\z@\@tempcntb\m@ne\def\@citea{}\@cite{\@for\@citeb:=#2\do
    {\@ifundefined
       {b@\@citeb}{\@citeo\@tempcntb\m@ne\@citea
        \def\@citea{,\penalty\@m\ }{\bf ?}\@warning
       {Citation `\@citeb' on page \thepage \space undefined}}%
    {\setbox\z@\hbox{\global\@tempcntc0\csname
b@\@citeb\endcsname\relax}%
     \ifnum\@tempcntc=\z@ \@citeo\@tempcntb\m@ne
       \@citea\def\@citea{,\penalty\@m}
       \hbox{\csname b@\@citeb\endcsname}%
     \else
      \advance\@tempcntb\@ne
      \ifnum\@tempcntb=\@tempcntc
      \else\advance\@tempcntb\m@ne\@citeo
      \@tempcnta\@tempcntc\@tempcntb\@tempcntc\fi\fi}}\@citeo}{#1}}

\def\@citeo{\ifnum\@tempcnta>\@tempcntb\else\@citea
  \def\@citea{,\penalty\@m}%
  \ifnum\@tempcnta=\@tempcntb\the\@tempcnta\else
   {\advance\@tempcnta\@ne\ifnum\@tempcnta=\@tempcntb \else
\def\@citea{--}\fi
    \advance\@tempcnta\m@ne\the\@tempcnta\@citea\the\@tempcntb}\fi\fi}

\def\nl{\nonumber\\}

\newcommand{\gsim}
{\mathrel{\raisebox{-.3em}{$\stackrel{\displaystyle >}{\sim}$}}}
\def\asymp#1%
{\mathrel{\raisebox{-.4em}{$\widetilde{\scriptstyle #1}$}}}

\def\Nequal#1%
{\mathrel{\raisebox{-.5em}{$\stackrel{=}{\scriptstyle\rm#1}$}}}
\newcommand{\dsl}[1]{\not \hspace{-0.7mm}#1}
\def\dsl{\mathpalette\make@slash}
\def\make@slash#1#2{\setbox\z@\hbox{$#1#2$}%
  \hbox to 0pt{\hss$#1/$\hss\kern-\wd0}\box0}

\def\beq{\begin{equation}}
\def\eeq{\end{equation}}
\def\beqar{\begin{eqnarray}}
\def\eeqar{\end{eqnarray}}
\def\barr#1{\begin{array}{#1}}
\def\earr{\end{array}}
\def\bfi{\begin{figure}}
\def\efi{\end{figure}}
\def\btab{\begin{table}}
\def\etab{\end{table}}
\def\bce{\begin{center}}
\def\ece{\end{center}}
\def\nn{\nonumber}
\def\disp{\displaystyle}
\def\text{\textstyle}

\def\arraystretch{1.2}

\def\al{\alpha}

\def\Ga{\Gamma}
\def\ga{\gamma}
\def\de{\delta}

\def\la{\lambda}

\def\si{\sigma}

\def\refeq#1{\mbox{(\ref{#1})}}
\def\refeqs#1{\mbox{(\ref{#1})}}
\def\refeqf#1{\mbox{(\ref{#1})}}
\def\reffi#1{\mbox{Figure~\ref{#1}}}

\def\refta#1{\mbox{Table~\ref{#1}}}

\def\refse#1{\mbox{Section~\ref{#1}}}
\def\refses#1{\mbox{Sections~\ref{#1}}}

\def\citere#1{\mbox{Ref.~\cite{#1}}}
\def\citeres#1{\mbox{Refs.~\cite{#1}}}

\newcommand{\TeV}{\unskip\,\mathrm{TeV}}
\newcommand{\GeV}{\unskip\,\mathrm{GeV}}

\newcommand{\fb}{\unskip\,\mathrm{fb}}

\newcommand{\ri}{{\mathrm{i}}}
\newcommand{\rd}{{\mathrm{d}}}

\newcommand{\Oa}{\mathswitch{{\cal{O}}(\alpha)}}

\newcommand{\M}{{\cal{M}}}

\def\mathswitchr#1{\relax\ifmmode{\mathrm{#1}}\else$\mathrm{#1}$\fi}

\newcommand{\PV}{\mathswitchr V}
\newcommand{\PW}{\mathswitchr W}
\newcommand{\Pw}{\mathswitchr w}
\newcommand{\PZ}{\mathswitchr Z}

\newcommand{\Pg}{\mathswitchr g}

\newcommand{\Pe}{\mathswitchr e}
\newcommand{\Pne}{\mathswitch \nu_{\mathrm{e}}}
\newcommand{\Pnebar}{\mathswitch \bar\nu_{\mathrm{e}}}
\newcommand{\Pd}{\mathswitchr d}
\newcommand{\Pdbar}{\bar{\mathswitchr d}}
\newcommand{\Pu}{\mathswitchr u}
\newcommand{\Pubar}{\bar{\mathswitchr u}}
\newcommand{\Ps}{\mathswitchr s}
\newcommand{\Psbar}{\bar{\mathswitchr s}}
\newcommand{\Pc}{\mathswitchr c}
\newcommand{\Pcbar}{\bar{\mathswitchr c}}

\newcommand{\Pep}{\mathswitchr {e^+}}
\newcommand{\Pem}{\mathswitchr {e^-}}
\newcommand{\PWp}{\mathswitchr {W^+}}
\newcommand{\PWm}{\mathswitchr {W^-}}

\def\mathswitch#1{\relax\ifmmode#1\else$#1$\fi}

\newcommand{\MV}{\mathswitch {M_\PV}}
\newcommand{\MW}{\mathswitch {M_\PW}}

\newcommand{\MZ}{\mathswitch {M_\PZ}}

\newcommand{\GW}{\Gamma_{\PW}}
\newcommand{\GZ}{\Gamma_{\PZ}}
\newcommand{\GV}{\Gamma_{\PV}}

\newcommand{\sw}{\mathswitch {s_\Pw}}
\newcommand{\cw}{\mathswitch {c_\Pw}}

\newcommand{\GF}{\mathswitch {G_\mu}}

\def\solid{\raise.9mm\hbox{\protect\rule{1.1cm}{.2mm}}}
\def\dash{\raise.9mm\hbox{\protect\rule{2mm}{.2mm}}\hspace*{1mm}}

\def\ie{i.e.\ }

\def\cf{cf.\ }

\renewcommand{\min}{{\mathrm{min}}}
\renewcommand{\max}{{\mathrm{max}}}

\newcommand{\U}{\mathrm{U}}
\newcommand{\SU}{\mathrm{SU}}

\def\Re{\mathop{\mathrm{Re}}\nolimits}

\newcommand{\eeffff}{\Pep\Pem\to 4f}
\newcommand{\eeffffg}{\eeffff\ga}

\newcommand{\spab}{\langle p_a p_b \rangle}
\newcommand{\spac}{\langle p_a p_c \rangle}
\newcommand{\spad}{\langle p_a p_d \rangle}
\newcommand{\spae}{\langle p_a p_e \rangle}

\newcommand{\spbd}{\langle p_b p_d \rangle}
\newcommand{\spbe}{\langle p_b p_e \rangle}
\newcommand{\spbf}{\langle p_b p_f \rangle}
\newcommand{\spcd}{\langle p_c p_d \rangle}
\newcommand{\spce}{\langle p_c p_e \rangle}

\newcommand{\spdf}{\langle p_d p_f \rangle}
\newcommand{\spef}{\langle p_e p_f \rangle}
\newcommand{\spak}{\langle p_a k \rangle}
\newcommand{\spbk}{\langle p_b k \rangle}
\newcommand{\spck}{\langle p_c k \rangle}
\newcommand{\spdk}{\langle p_d k \rangle}
\newcommand{\spek}{\langle p_e k \rangle}
\newcommand{\spfk}{\langle p_f k \rangle}

\newcommand{\cspad}{\spad^*}

\newcommand{\cspbd}{\spbd^*}

\newcommand{\cspbf}{\spbf^*}
\newcommand{\cspcd}{\spcd^*}

\newcommand{\cspdf}{\spdf^*}

\newcommand{\cspbk}{\spbk^*}

\newcommand{\cspdk}{\spdk^*}

\newcommand{\cspfk}{\spfk^*}

\def\draftdate{\relax}
\def\mda{\relax}
\def\mua{\relax}
\def\mla{\relax}
\def\draft{
\def\thtystars{******************************}
\def\sixtystars{\thtystars\thtystars}
\typeout{}
\typeout{\sixtystars**}
\typeout{* Draft mode!
         For final version remove \protect\draft\space in source file *}
\typeout{\sixtystars**}
\typeout{}
\def\draftdate{\today}
\def\mua{\marginpar[\boldmath\hfil$\uparrow$]%
                   {\boldmath$\uparrow$\hfil}%
                    \typeout{marginpar: $\uparrow$}\ignorespaces}
\def\mda{\marginpar[\boldmath\hfil$\downarrow$]%
                   {\boldmath$\downarrow$\hfil}%
                    \typeout{marginpar: $\downarrow$}\ignorespaces}
\def\mla{\marginpar[\boldmath\hfil$\rightarrow$]%
                   {\boldmath$\leftarrow $\hfil}%
                    \typeout{marginpar: $\leftrightarrow$}\ignorespaces}
\def\Mua{\marginpar[\boldmath\hfil$\Uparrow$]%
                   {\boldmath$\Uparrow$\hfil}%
                    \typeout{marginpar: $\uparrow$}\ignorespaces}
\def\Mda{\marginpar[\boldmath\hfil$\Downarrow$]%
                   {\boldmath$\Downarrow$\hfil}%
                    \typeout{marginpar: $\downarrow$}\ignorespaces}
\def\Mla{\marginpar[\boldmath\hfil$\Rightarrow$]%
                   {\boldmath$\Leftarrow $\hfil}%
                    \typeout{marginpar: $\leftrightarrow$}\ignorespaces}
\overfullrule 5pt
\oddsidemargin -15mm
\marginparwidth 29mm
}

\def\stars{\strut\leaders\hbox{*}\hfill\strut}
\def\starline{\hfil\strut\hfil\hbox to \textwidth {\stars}\hfil}

\begin{document}
\thispagestyle{empty}
\def\thefootnote{\fnsymbol{footnote}}
\setcounter{footnote}{1}
\null
\draftdate\hfill BI-TP 99/10 \\
\strut\hfill PSI-PR-99-12\\
\strut\hfill hep-ph/9904472
\vfill
\begin{center}
{\Large \bf\boldmath
Predictions for all processes $\Pep\Pem\to4 \mbox{ fermions} + \ga$%
\par} \vskip 2.5em
\vspace{1cm}

{\large
{\sc A.\ Denner$^1$, S.\ Dittmaier$^2$, M. Roth$^{1,3}$ and 
D.\ Wackeroth$^1$} } \\[1cm]

$^1$ {\it Paul-Scherrer-Institut, W\"urenlingen und Villigen\\
CH--5232 Villigen PSI, Switzerland} \\[0.5cm]

$^2$ {\it Theoretische Physik, Universit\"at Bielefeld \\
D-33615 Bielefeld, Germany}
\\[0.5cm]

$^3$ {\it Institut f\"ur Theoretische Physik, ETH-H\"onggerberg\\
CH--8093 Z\"urich, Switzerland}
\par \vskip 1em
\end{center}\par
\vskip 2cm
{\bf Abstract:} \par
The complete matrix elements for $\eeffff$ and $\eeffffg$ are
calculated in the Electroweak Standard Model for 
polarized massless fermions. The matrix elements for all final states are
reduced to a few compact generic functions. Monte Carlo generators for
$\eeffff$ and $\eeffffg$ are constructed.
We compare different treatments of
the finite widths of the electroweak gauge bosons; in particular, we
include a scheme with a complex gauge-boson 
mass that obeys all Ward identities.
The detailed discussion of numerical results comprises integrated cross
sections as well as photon-energy distributions for all different final
states.
\par
\vskip 1cm
\noindent
April 1999
\null
\setcounter{page}{0}
\clearpage
\def\thefootnote{\arabic{footnote}}
\setcounter{footnote}{0}

\section{Introduction}
\label{se:intro}

When exceeding the \PW-pair production threshold, the LEP collider
started a new era in the verification of the Electroweak Standard
Model (SM): the study of the properties of the \PW~boson and of its
interactions.
While the most important process at
LEP2 in this respect is certainly $\Pep\Pem\to\PWp\PWm\to4f$, many
other reactions have now become accessible. Besides the  
4-fermion-production processes, including single \PW-boson production,
single \PZ-boson production, or \PZ-boson-pair production, LEP2 
and especially a future linear collider allow us
to investigate another class of processes, namely
$\eeffffg$.

The physical interest in the processes $\eeffffg$ is
twofold. First of all, they are an important building block for the
radiative corrections to $\Pep\Pem\to4f$, and their effect must be
taken into account in order to get precise predictions for the
observables that are used for the measurement of the \PW-boson mass
and the 
triple-gauge-boson couplings. 
On the other hand, those processes themselves involve interesting
physics. They include, in particular, triple-gauge-boson-production
processes 
such as $\PWp\PWm\ga$, $\PZ\PZ\ga$, or $\PZ\ga\ga$ production
and can therefore be used to obtain information on the quartic
gauge-boson couplings $\ga\ga WW$, $\ga ZWW$, and $\ga\ga ZZ$.
While only a few events of this kind are expected at LEP2,
these studies can be performed in more detail at future linear
$\Pep\Pem$ colliders \cite{Be92}. 

Important contributions to $\eeffff$ and $\eeffffg$ arise from
subprocesses with two resonant gauge bosons. These
subprocesses do not only contain interesting physics, such as the
quartic gauge-boson self-interactions, but also dominate the cross
sections in those regions of phase space where the invariant masses of
certain combinations of final-state particles are close to the mass of
the corresponding nearly resonant
gauge bosons. 
Therefore, it is sometimes a reasonable approach to consider only
the resonant contributions.
In the
most naive approximation, the produced gauge bosons are treated as
stable particles.  Most of the existing calculations for
triple-gauge-boson production have been performed in this way, such as
many calculations for $\Pep\Pem\to\PWp\PWm\ga$ \cite{WWg}.

In an improved approach, the so-called pole expansion
\cite{polescheme,Ae94}, the resonances are treated exactly, \ie the
decay of the produced gauge bosons is taken into account, but the
matrix elements are expanded about the poles of the resonances. Once
the non-resonant diagrams have been left out, the expansion is in fact
mandatory in order to retain gauge invariance. If only the leading
terms in the expansion are kept, this approach is known as the pole
approximation, or double-pole approximation in the presence of two
resonances. 
The accuracy of the (double-)pole approximation is, at best, of the
order of $\GV/\MV$, where $\MV$ and $\GV$ are the mass and the width
of the relevant gauge bosons, and thus typically at the level of
several per cent. Consequently, this approach is only reasonable if the
experimental accuracy is correspondingly low. 
The error estimate of $\GV/\MV$ is too optimistic in situations in which
scales of order $\GV$, or smaller, are involved. This is, in particular,
the case in threshold regions or if photons with energies of the order 
of $\GV$ are emitted from the resonant gauge bosons.
On the other hand, the quality of the pole approximation can be
improved by applying appropriate invariant-mass cuts.

Note that the double-pole approximation is particularly well-suited
for the calculation of the radiative corrections to
gauge-boson pair production, since the error resulting from
the double-pole approximation is suppressed by an additional factor
$\al/\pi$ in this case. In this approximation the
corrections can be classified into factorizable and non-factorizable
corrections \cite{Ae94,wwrev}. The virtual factorizable corrections
can be composed of the known corrections to 
on-shell \PW-pair
production \cite{rcwprod} and on-shell \PW~decay \cite{rcwdecay}, and the
non-factorizable corrections have recently been calculated \cite{nfc}.
In contrast to the virtual factorizable corrections, 
the real factorizable corrections cannot be simply taken over from the
on-shell processes. In the case of real photon
emission the definition of the double-pole approximation is
non-trivial if the energy of the final-state photon is of the order of
$\GV$, since the resonances before and after photon emission are not
well separated in phase space.
A possible double-pole approximation of the $\Oa$
corrections to four-lepton production has been discussed in \citere{be98b}.

In our calculation of $\eeffffg$ we 
do not use the double-pole approximation since the full calculation is
feasable with reasonable effort. Unlike the double-pole
approximation, the full calculation is valid for arbitrary processes
in the set $\eeffffg$. Moreover, when using the exact results for the
real corrections in an $\Oa$ calculation of 4 fermion-production
processes, the reliability of possible approximations can be tested.


Some results for $\eeffffg$ with an observable photon already exist in
the literature. In \citeres{Ae91,Ae91a} the contributions to the
matrix elements involving two resonant \PW~bosons have been calculated
and implemented into a Monte Carlo generator.  This generator has been
extended to include collinear bremsstrahlung \cite{vO94} and used to
discuss the effect of hard photons at LEP2 \cite{vO96}. The complete
cross section for the process $\Pep\Pem\to\Pu\,\Pdbar\,\Pem\Pnebar\ga$ has
been discussed in \citere{Fu94}.  In \citere{Ca97}, the complete
matrix elements for the processes $\eeffffg$ have been calculated
using an iterative numerical algorithm without referring to Feynman
diagrams. We are, however, interested in explicit analytical results
on the amplitudes for various reasons.  In particular, we want to have
full control over the implementation of the finite width of the
virtual vector bosons and to select single diagrams, such as the
doubly-resonant ones.
So far no results for $\eeffffg$ with $\Pep\Pem$ pairs in the final
state have been published.

In order to perform the calculation as efficient as possible we have
reduced all processes to a small number of generic contributions. For
$\eeffff$, the calculation is similar to the one in \citere{Be94},
and the generic contributions correspond to individual Feynman diagrams.
In the case of $\eeffffg$ we have combined groups of diagrams 
in such a way that the
resulting generic contributions can be classified in the same way as
those for $\eeffff$. As a consequence, the generic contributions are
individually gauge-invariant with respect to the external photon. The
number and the complexity of diagrams in the generic contributions for
$\eeffffg$ 
has been reduced by using a non-linear gauge-fixing condition
for the W-boson field
\cite{nlgauge}. In this way, many cancellations between diagrams are
avoided, without any 
further algebraic manipulations. Finally, for the helicity amplitudes
corresponding to the generic contributions concise results 
have been obtained by using the 
Weyl--van~der~Waerden formalism (see
\citere{wvdw} and references therein).

After the matrix elements have been calculated, the finite widths 
of the resonant particles have to be introduced. We have done this in 
different
ways and compared the different treatments for $\eeffff$ and
$\eeffffg$. In particular, we have discussed a ``complex-mass
scheme'', which preserves all Ward identities and is still rather
simple to apply. 

The matrix elements to $\eeffff$ and $\eeffffg$
exhibit a complex peaking behaviour owing to propagators
of massless particles and Breit--Wigner resonances,
so that the integration over the
8- and 11-dimensional phase 
spaces, respectively, is not straightforward.
In order to obtain numerically stable results,
we adopt the multi-channel integration
method \cite{Be94,Multichannel} and 
reduce  the Monte Carlo error by the adaptive weight optimization
procedure described in \citere{Kl94}.
In the multi-channel approach, 
we define a suitable mapping of random numbers into phase space variables
for each arising propagator structure.
These variables are generated according to distributions that
approximate this specific peaking behavior of the integrand.
For $\eeffff$ and $\eeffffg$ we identify up to 128 and 928 channels,
respectively, which necessitates an efficient and generic procedure
for the phase-space generation.  We wrote two independent Monte Carlo
programs following the general strategy outlined above. 
They differ in the realization of a generic procedure for the construction
of the phase-space generators.

The paper is organized as follows: in \refse{se:anres} we describe the
calculation of the helicity amplitudes for 
$\eeffff$ and $\eeffffg$ and list the complete results. The Monte Carlo 
programs are
described in \refse{se:MC}. In \refse{se:numres} the numerical results
are discussed, and \refse{se:sum} 
contains a summary and an outlook.

\section{Analytical results}
\label{se:anres}
\subsection{Notation and conventions}
\label{se:not&con}

We consider reactions of the types
\beqar\label{eq:eeffff}
\Pep(p_+,\si_+)+\Pem(p_-,\si_-) &\to& 
f_1(k_1,\si_1)+\bar f_2(k_2,\si_2)+f_3(k_3,\si_3)+\bar f_4(k_4,\si_4),
\qquad\\
\Pep(p_+,\si_+)+\Pem(p_-,\si_-) &\to& 
f_1(k_1,\si_1)+\bar f_2(k_2,\si_2)+f_3(k_3,\si_3)+\bar f_4(k_4,\si_4)
+\gamma(k_5,\lambda).\qquad
\label{eq:eeffffg}
\eeqar
The arguments label the momenta $p_\pm$, $k_i$ and helicities
$\si_i=\pm1/2$, $\la=\pm1$ of the corresponding particles. We often
use only the signs to denote the helicities. The fermion masses are
neglected everywhere.

For the Feynman rules we use the conventions of \citere{sm}. In
particular, all fields and momenta are incoming. It is
convenient to use a non-linear gauge-fixing term \cite{nlgauge}
of the form
\beqar\label{eq:nlgauge}
{\cal L}_{\mathrm{fix}} &=& -\left| \partial^\mu W^+_\mu + \ri e
                (A^\mu - \frac{\cw}{\sw} Z^\mu) W^+_\mu
               -\ri \MW \phi^+ \right|^2                        \nn\\*
            & &{}-  \frac{1}{2} (\partial^\mu Z_\mu - \MZ \chi)^2
              - \frac{1}{2} (\partial^\mu A_\mu)^2 \;,
\eeqar
where $\phi^\pm$ and $\chi$ are the would-be Goldstone bosons of the 
$W^\pm$ and $Z$ fields, respectively.
With this choice, the $\phi^\pm W^\mp A$ vertices vanish, and the
bosonic couplings that are relevant for $\eeffffg$ read
\beqar
\setlength{\unitlength}{1pt}
\label{fr:vvv}
\barr{l}
\begin{picture}(90,80)(-50,-38)
\Text(-45,3)[lb]{$V_{\mu},k_{V}$}
\Text(35,27)[rb]{$W^+_{\nu},k_{+}$}
\Text(35,-27)[rt] {$W^-_{\rho},k_{-}$}
\Vertex(0,0){2}
\Photon(0,0)(35,25){2}{3.5}
\Photon(0,0)(35,-25){2}{3.5}
\Photon(0,0)(-45,0){2}{3.5}
\end{picture} 
\earr
&&\barr{l}
=  -\ri e g_{VWW}\left[
g_{\nu\rho}(k_- -k_+)_\mu-2g_{\mu\nu}k_{V,\rho}+2g_{\mu\rho}k_{V,\nu} \right],
\nn\\
\earr\\
\label{fr:vvvv}
\barr{l}
\begin{picture}(90,80)(-50,-36)
\Text(-35,27)[lb]{$A_{\mu}$}
\Text(-35,-27)[lt]{$V_{\nu}$}
\Text(35,27)[rb]{$W^+_{\rho}$}
\Text(35,-27)[rt]{$W^-_{\sigma}$}
\Vertex(0,0){2}
\Photon(0,0)(35,25){2}{3.5}
\Photon(0,0)(35,-25){2}{3.5}
\Photon(0,0)(-35,25){2}{3.5}
\Photon(0,0)(-35,-25){2}{3.5}
\end{picture} 
\earr
&&\barr{l}
= -2\ri e g_{VWW} \, g_{\mu\nu}g_{\rho\sigma},
\earr
\eeqar
with $V=A,Z$, and the coupling factors
\beq
g_{AWW} = 1, \qquad g_{ZWW} = -\frac{\cw}{\sw}.
\eeq
Note that the gauge-boson propagators have the same simple form as in
the 't~Hooft--Feynman gauge, i.e.\ they are proportional to the metric
tensor $g_{\mu\nu}$.
This gauge choice eliminates some diagrams and simplifies others owing
to the simpler structure of the photon--gauge-boson couplings.

The vector-boson--fermion--fermion couplings have the usual form
\beq
\label{fr:Vff}
\barr{l}
\begin{picture}(90,80)(-50,-36)
\Text(-45,5)[lb]{$V_{\mu}$}
\Text(35,27)[rb]{$\bar{f}_{i}$}
\Text(35,-27)[rt]{$f_{j}$}
\Vertex(0,0){2}
\ArrowLine(0,0)(35,25)
\ArrowLine(35,-25)(0,0)
\Photon(0,0)(-45,0){2}{3.5}
\end{picture} 
\earr
\barr{l}
= \disp\ri e \gamma_\mu \sum_\si g^\sigma_{V\bar f_i f_j}\omega_\sigma,
\earr
\eeq
where
$\omega_\pm=(1\pm\gamma_5)/2$.
The corresponding coupling factors read 
\beq
g^\sigma_{A\bar f_i f_i} = -Q_i, \qquad
g^\sigma_{Z\bar f_i f_i} = 
-\frac{\sw}{\cw}Q_i+\frac{I^3_{{\mathrm{w}},i}}{\cw\sw}\delta_{\sigma-}, 
\qquad
g^\sigma_{W\bar f_i f'_i} = \frac{1}{\sqrt{2}\sw}\delta_{\sigma-},
\eeq
where $Q_i$ and $I^3_{{\mathrm{w}},i}=\pm{1/2}$ denote the
relative charge and the weak isospin of the fermion $f_i$,
respectively, and $f'_i$ is the weak-isospin partner of $f_i$.
The colour factor of a fermion $f_i$ is denoted by 
$N^{\mathrm{c}}_{f_i}$, i.e.\
$N^{\mathrm{c}}_{\mathrm{lepton}}=1$ and
$N^{\mathrm{c}}_{\mathrm{quark}}=3$.
 
\subsection{Classification of final states for $\eeffff$}
\label{se:pclass}

The final states for $\eeffff$ have 
already been classified in
\citeres{Be94,Ba94,CERN9601mcgen}. We introduce a classification that
is very close to the one of \citeres{Ba94,CERN9601mcgen}.  It is based
on the production
mechanism, i.e.\ whether the 
reactions proceed via charged-current (CC), or
neutral-current (NC) interactions, or via both interaction types.  The
classification can be performed by considering the quantum numbers of
the final-state fermion pairs.  In the following, $f$ and $F$ denote
different fermions ($f\ne F$) that are neither electrons nor electron
neutrinos
($f,F \ne \Pem,\nu_\Pe$), and their weak-isospin partners are denoted
by $f'$ and $F'$, respectively. We find the following 11 classes of
processes (in parenthesis the corresponding classification of
\citere{CERN9601mcgen} is given):
\renewcommand{\labelenumi}{(\roman{enumi})}
\renewcommand{\labelenumii}{(\alph{enumii})}
\newcommand{\ientry}[2]{\mbox{\rlap{#1}\hspace*{5cm}#2}}
\begin{enumerate}
{\samepage
\item CC reactions:
\begin{enumerate}
\item \ientry{$\Pep\Pem \to f \bar f' F \bar F'$,}%
             {({\em CC11} family),} 
\item \ientry{$\Pep\Pem \to \nu_\Pe \Pep f \bar f'$,}%
             {({\em CC20} family),}
\item \ientry{$\Pep\Pem \to f \bar f' \Pem \bar\nu_\Pe$,}%
             {({\em CC20} family),}
\end{enumerate}
}
\item NC reactions:
\begin{enumerate}
\item \ientry{$\Pep\Pem \to f \bar f F \bar F$,}%
             {({\em NC32} family),}
\item \ientry{$\Pep\Pem \to f \bar f f \bar f$,}%
             {({\em NC4$\cdot$16} family),}
\item \ientry{$\Pep\Pem \to \Pem \Pep f \bar f$,}%
             {({\em NC48} family),}
\item \ientry{$\Pep\Pem \to \Pem \Pep \Pem \Pep$,}%
             {({\em NC4$\cdot$36} family),}
\end{enumerate}
\item Mixed CC/NC reactions:
\begin{enumerate}
\item \ientry{$\Pep\Pem \to f \bar f f' \bar f'$,}%
             {({\em mix43} family),}
\item \ientry{$\Pep\Pem \to \nu_\Pe \bar\nu_\Pe f \bar f$,}%
             {({\em NC21} family),}
\item \ientry{$\Pep\Pem \to \nu_\Pe \bar\nu_\Pe \nu_\Pe \bar\nu_\Pe$,}%
             {({\em NC4$\cdot$9} family),}
\item \ientry{$\Pep\Pem \to \nu_\Pe \bar\nu_\Pe \Pem \Pep$,}%
             {({\em mix56} family).}
\end{enumerate}
\end{enumerate}
The radiation of an additional photon does not change this
classification.

\subsection{Generic diagrams and amplitudes for \boldmath{$\Pep\Pem\to 4f$} }
\label{se:ee4f}

In order to explain and to illustrate our generic approach we first
list the results for $\eeffff$. All these processes can be composed
from only two generic diagrams, the abelian and non-abelian diagrams
shown in \reffi{ee4fdiags}. All external fermions $f_{a,\dots,f}$ are
assumed to be incoming, and the momenta and helicities are denoted by
$p_{a,\dots,f}$ and $\si_{a,\dots,f}$, respectively.
\bfi
\begin{center}
\setlength{\unitlength}{1pt}
\begin{picture}(420,150)(0,-20)
\Text(0,120)[lb]{a) abelian graph}
\Text(210,120)[lb]{b) non-abelian graph}
\put(20,-8){
\begin{picture}(150,100)(0,0)
\ArrowLine(35,70)( 5, 95)
\ArrowLine( 5, 5)(35, 30)
\ArrowLine(35,30)(35,70)
\Photon(35,30)(90,20){2}{6}
\Photon(35,70)(90,80){-2}{6}
\Vertex(35,70){2.0}
\Vertex(35,30){2.0}
\Vertex(90,80){2.0}
\Vertex(90,20){2.0}
\ArrowLine(90,80)(120, 95)
\ArrowLine(120,65)(90,80)
\ArrowLine(120, 5)( 90,20)
\ArrowLine( 90,20)(120,35)
\put(55,82){$V_1$}
\put(55,10){$V_2$}
\put( 20,50){$f_g$}
\put(-15,110){$\bar f_a(p_a,\si_a)$}
\put(-15,-10){$f_b(p_b,\si_b)$}
\put(125,90){$\bar f_c(p_c,\si_c)$}
\put(125,65){$f_d(p_d,\si_d)$}
\put(125,30){$\bar f_e(p_e,\si_e)$}
\put(125, 5){$f_f(p_f,\si_f)$}
\end{picture}
}
\put(245,-8){
\begin{picture}(150,100)(0,0)
\ArrowLine( 15,50)(-15, 65)
\ArrowLine(-15,35)( 15, 50)
\Photon(15,50)(60,50){2}{5}
\Photon(60,50)(90,20){-2}{5}
\Photon(60,50)(90,80){2}{5}
\Vertex(15,50){2.0}
\Vertex(60,50){2.0}
\Vertex(90,80){2.0}
\Vertex(90,20){2.0}
\ArrowLine(90,80)(120, 95)
\ArrowLine(120,65)(90,80)
\ArrowLine(120, 5)( 90,20)
\ArrowLine( 90,20)(120,35)
\put(30,58){$V$}
\put(62,70){$W$}
\put(62,18){$W$}
\put(-35,75){$\bar f_a(p_a,\si_a)$}
\put(-35,20){$f_b(p_b,\si_a)$}
\put(125,90){$\bar f_c(p_c,\si_c)$}
\put(125,65){$f_d(p_d,\si_a)$}
\put(125,30){$\bar f_e(p_e,\si_e)$}
\put(125, 5){$f_f(p_f,\si_f)$}
\end{picture}
}
\end{picture}
\end{center}
\caption{Generic diagrams for $\Pep\Pem\to 4f$}
\label{ee4fdiags}
\efi
The helicity amplitudes of these diagrams are calculated within the
Weyl--van~der~Waerden (WvdW) formalism following the conventions of 
\citere{wvdw} (see also references therein). 

\subsubsection{Leptonic and semi-leptonic final states}

We first treat purely leptonic and semi-leptonic final states. In this
case, none of the gauge bosons in the generic graphs of \reffi{ee4fdiags}
can be a gluon, and the colour structure trivially leads to a global
factor $N^{\mathrm{c}}_{f_1}N^{\mathrm{c}}_{f_3}$, which is equal to 1
or 3, after summing the squared amplitude over the colour degrees of 
freedom. The results for the generic amplitudes are
\beqar\label{genfun4fNC}
&& \hspace*{-3em}
{\cal M}^{\si_a,\si_b,\si_c,\si_d,\si_e,\si_f}_{V_1 V_2}
(p_a,p_b,p_c,p_d,p_e,p_f) 
\nn\\*
\hspace*{2em}
&=& -4e^4 \de_{\si_a,-\si_b} \de_{\si_c,-\si_d} \de_{\si_e,-\si_f} \, 
g^{\si_b}_{V_1\bar f_a f_g} g^{\si_b}_{V_2\bar f_g f_b}
g^{\si_d}_{V_1\bar f_c f_d} g^{\si_f}_{V_2\bar f_e f_f}
\nn\\ && {} \times
\frac{P_{V_1}(p_c+p_d) P_{V_2}(p_e+p_f)}{(p_b+p_e+p_f)^2} \,
A^{\si_a,\si_c,\si_e}_2(p_a,p_b,p_c,p_d,p_e,p_f),
\hspace*{3em}
\\[1em] 
&& \hspace*{-3em}
{\cal M}^{\si_a,\si_b,\si_c,\si_d,\si_e,\si_f}_{VWW}
(p_a,p_b,p_c,p_d,p_e,p_f) 
\nn\\*
&=& -4e^4 \de_{\si_a,-\si_b} \de_{\si_c,+} \de_{\si_d,-} 
\de_{\si_e,+} \de_{\si_f,-} \, 
(Q_c-Q_d)g_{VWW} g^{\si_b}_{V\bar f_a f_b} 
g^{-}_{W\bar f_c f_d} g^{-}_{W\bar f_e f_f}
\nn\\ && {} \times
P_V(p_a+p_b) P_W(p_c+p_d) P_W(p_e+p_f) \,
A^{\si_a}_3(p_a,p_b,p_c,p_d,p_e,p_f),
\label{genfun4fCC}
\eeqar
where 
the vector-boson propagators are abbreviated by
\beq
P_V(p) = \frac{1}{p^2-\MV^2}, \qquad V=A,Z,W,g, \qquad M_A = M_g = 0.
\eeq
(The case of the gluon is included for later convenience.)
The auxiliary functions $A^{\si_a,\si_c,\si_e}_2$ and $A^{\si_a}_3$ are
expressed in terms of WvdW spinor products,
\beqar
A^{{+}{+}{+}}_2(p_a,p_b,p_c,p_d,p_e,p_f) &=& 
\spac\cspbf(\cspbd\spbe+\cspdf\spef),
\nn\\
A^{{+}{+}{-}}_2(p_a,p_b,p_c,p_d,p_e,p_f) &=& 
A^{{+}{+}{+}}_2(p_a,p_b,p_c,p_d,p_f,p_e),
\nn\\
A^{{+}{-}{+}}_2(p_a,p_b,p_c,p_d,p_e,p_f) &=& 
A^{{+}{+}{+}}_2(p_a,p_b,p_d,p_c,p_e,p_f),
\nn\\
A^{{+}{-}{-}}_2(p_a,p_b,p_c,p_d,p_e,p_f) &=& 
A^{{+}{+}{+}}_2(p_a,p_b,p_d,p_c,p_f,p_e),
\nn\\
A^{{-},\si_c,\si_d}_2(p_a,p_b,p_c,p_d,p_e,p_f) &=& 
\left(A^{{+},-\si_c,-\si_d}_2(p_a,p_b,p_c,p_d,p_e,p_f)\right)^*,
\\[1em]
A^{+}_3(p_a,p_b,p_c,p_d,p_e,p_f) &=& \phantom{{}+{}}
  \cspbd\cspbf\spab\spce + \cspbd\cspdf\spae\spcd
\nn\\
&& {}
+ \cspbf\cspdf\spac\spef,
\nn\\
A^{-}_3(p_a,p_b,p_c,p_d,p_e,p_f) &=&
A^{+}_3(p_b,p_a,p_c,p_d,p_e,p_f).
\eeqar
The spinor products are defined by
\beq
\langle pq\rangle=\epsilon^{AB}p_A q_B
=2\sqrt{p_0 q_0} \,\Biggl[
{\mathrm{e}}^{-\ri\phi_p}\cos\frac{\theta_p}{2}\sin\frac{\theta_q}{2}
-{\mathrm{e}}^{-\ri\phi_q}\cos\frac{\theta_q}{2}\sin\frac{\theta_p}{2}
\Biggr], 
\eeq
where $p_A$, $q_A$ are the associated momentum spinors for the momenta
\beqar
p^\mu&=&p_0(1,\sin\theta_p\cos\phi_p,\sin\theta_p\sin\phi_p,\cos\theta_p),\nl
q^\mu&=&p_0(1,\sin\theta_q\cos\phi_q,\sin\theta_q\sin\phi_q,\cos\theta_q).
\eeqar 

Incoming fermions are turned into outgoing ones by crossing, which
is performed by inverting the corresponding fermion momenta and
helicities.
If the generic functions are called with negative momenta $-p$, $-q$, it
is understood that only the complex conjugate spinor products get the
corresponding sign change. We illustrate this by 
simple examples:
\beq
\begin{array}[b]{rllll}
A(p,q)   &= \langle pq \rangle
&=\phantom{-}A(p,-q)  
&=\phantom{-}A(-p,q)  
&=A(-p,-q), \\
B(p,q)   &= \langle pq \rangle^*
&=-B(p,-q)  
&=-B(-p,q)  
&=B(-p,-q). 
\end{array}
\eeq
We have checked 
the results for the generic diagrams against those of \citere{Be94}
and found 
agreement.

Using the results for the generic diagrams of \reffi{ee4fdiags}, the
helicity amplitudes for all possible processes involving 
six external fermions can be built up. 
It is convenient to construct first 
the amplitudes for the process types 
CC(a) and NC(a) (see \refse{se:pclass}) in terms of the generic functions
\refeq{genfun4fNC} and \refeq{genfun4fCC}, because these amplitudes 
are the basic subamplitudes of 
the other channels. The full amplitude for each
process type can be built up from those subamplitudes
by appropriate substitutions and linear combinations.

We first list the helicity amplitudes for the CC processes:
\beqar
\label{eq:ee4f_cca}
&& \hspace*{-4em}
{\cal M}^{\si_+,\si_-,\si_1,\si_2,\si_3,\si_4}_{\mathrm{CCa}}
(p_+,p_-,k_1,k_2,k_3,k_4) 
\nn\\ &=& 
{\cal M}^{\si_+,\si_-,-\si_1,-\si_2,-\si_3,-\si_4}_{WW}
(p_+,p_-,-k_1,-k_2,-k_3,-k_4)
\nn\\ && {}
+ \sum_{V=\gamma,Z} \Bigl[ \phantom{{}+{}}
{\cal M}^{\si_+,\si_-,-\si_1,-\si_2,-\si_3,-\si_4}_{VWW}
(p_+,p_-,-k_1,-k_2,-k_3,-k_4)
\nn\\ && \hphantom{{} + \sum_{V=\gamma,Z} \Bigl[}
+ {\cal M}^{-\si_1,-\si_2,\si_+,\si_-,-\si_3,-\si_4}_{VW}
(-k_1,-k_2,p_+,p_-,-k_3,-k_4)
\nn\\ && \hphantom{{} + \sum_{V=\gamma,Z} \Bigl[}
+ {\cal M}^{-\si_3,-\si_4,\si_+,\si_-,-\si_1,-\si_2}_{VW}
(-k_3,-k_4,p_+,p_-,-k_1,-k_2)
\nn\\ && \hphantom{{} + \sum_{V=\gamma,Z} \Bigl[}
+ {\cal M}^{-\si_1,-\si_2,-\si_3,-\si_4,\si_+,\si_-}_{WV}
(-k_1,-k_2,-k_3,-k_4,p_+,p_-)
\nn\\ && \hphantom{{} + \sum_{V=\gamma,Z} \Bigl[}
+ {\cal M}^{-\si_3,-\si_4,-\si_1,-\si_2,\si_+,\si_-}_{WV}
(-k_3,-k_4,-k_1,-k_2,p_+,p_-) \Bigr],
\\[1em]
&& \hspace*{-4em}
{\cal M}^{\si_+,\si_-,\si_1,\si_2,\si_3,\si_4}_{\mathrm{CCb}}
(p_+,p_-,k_1,k_2,k_3,k_4) 
\nn\\ &=& \phantom{{}+{}}
{\cal M}^{\si_+,\si_-,\si_1,\si_2,\si_3,\si_4}_{\mathrm{CCa}}
(p_+,p_-,k_1,k_2,k_3,k_4) 
\nn\\ && {}
- {\cal M}^{\si_+,-\si_2,\si_1,-\si_-,\si_3,\si_4}_{\mathrm{CCa}}
(p_+,-k_2,k_1,-p_-,k_3,k_4),
\\[1em]
&& \hspace*{-4em}
{\cal M}^{\si_+,\si_-,\si_1,\si_2,\si_3,\si_4}_{\mathrm{CCc}}
(p_+,p_-,k_1,k_2,k_3,k_4) 
\nn\\ &=& \phantom{{}+{}}
{\cal M}^{\si_+,\si_-,\si_1,\si_2,\si_3,\si_4}_{\mathrm{CCa}}
(p_+,p_-,k_1,k_2,k_3,k_4) 
\nn\\ && {}
- {\cal M}^{-\si_3,\si_-,\si_1,\si_2,-\si_+,\si_4}_{\mathrm{CCa}}
(-k_3,p_-,k_1,k_2,-p_+,k_4).
\eeqar
The ones for the NC processes are given by
\beqar
\label{eq:ee4f_nca}
&& \hspace*{-4em}
{\cal M}^{\si_+,\si_-,\si_1,\si_2,\si_3,\si_4}_{\mathrm{NCa}}
(p_+,p_-,k_1,k_2,k_3,k_4) 
\nn\\ &=& \sum_{V_1,V_2=\gamma,Z} \Bigl[ \phantom{{}+{}}
{\cal M}^{\si_+,\si_-,-\si_1,-\si_2,-\si_3,-\si_4}_{V_1 V_2}
(p_+,p_-,-k_1,-k_2,-k_3,-k_4)
\nn\\ && \hphantom{\sum_{V_1,V_2=\gamma,Z} \Bigl[}
+ {\cal M}^{\si_+,\si_-,-\si_3,-\si_4,-\si_1,-\si_2}_{V_1 V_2}
(p_+,p_-,-k_3,-k_4,-k_1,-k_2)
\nn\\ && \hphantom{\sum_{V_1,V_2=\gamma,Z} \Bigl[}
+ {\cal M}^{-\si_1,-\si_2,\si_+,\si_-,-\si_3,-\si_4}_{V_1 V_2}
(-k_1,-k_2,p_+,p_-,-k_3,-k_4)
\nn\\ && \hphantom{\sum_{V_1,V_2=\gamma,Z} \Bigl[}
+ {\cal M}^{-\si_3,-\si_4,\si_+,\si_-,-\si_1,-\si_2}_{V_1 V_2}
(-k_3,-k_4,p_+,p_-,-k_1,-k_2)
\nn\\ && \hphantom{\sum_{V_1,V_2=\gamma,Z} \Bigl[}
+ {\cal M}^{-\si_1,-\si_2,-\si_3,-\si_4,\si_+,\si_-}_{V_1 V_2}
(-k_1,-k_2,-k_3,-k_4,p_+,p_-)
\nn\\ && \hphantom{\sum_{V_1,V_2=\gamma,Z} \Bigl[}
+ {\cal M}^{-\si_3,-\si_4,-\si_1,-\si_2,\si_+,\si_-}_{V_1 V_2}
(-k_3,-k_4,-k_1,-k_2,p_+,p_-) \Bigl],
\\[1em]
&& \hspace*{-4em}
{\cal M}^{\si_+,\si_-,\si_1,\si_2,\si_3,\si_4}_{\mathrm{NCb}}
(p_+,p_-,k_1,k_2,k_3,k_4) 
\nn\\* &=& \phantom{{}+{}}
{\cal M}^{\si_+,\si_-,\si_1,\si_2,\si_3,\si_4}_{\mathrm{NCa}}
(p_+,p_-,k_1,k_2,k_3,k_4) 
\nn\\ && {}
- {\cal M}^{\si_+,\si_-,\si_3,\si_2,\si_1,\si_4}_{\mathrm{NCa}}
(p_+,p_-,k_3,k_2,k_1,k_4),
\\[1em]
&& \hspace*{-4em}
{\cal M}^{\si_+,\si_-,\si_1,\si_2,\si_3,\si_4}_{\mathrm{NCc}}
(p_+,p_-,k_1,k_2,k_3,k_4) 
\nn\\* &=& \phantom{{}+{}}
{\cal M}^{\si_+,\si_-,\si_1,\si_2,\si_3,\si_4}_{\mathrm{NCa}}
(p_+,p_-,k_1,k_2,k_3,k_4) 
\nn\\ && {}
- {\cal M}^{-\si_1,\si_-,-\si_+,\si_2,\si_3,\si_4}_{\mathrm{NCa}}
(-k_1,p_-,-p_+,k_2,k_3,k_4),
\\[1em]
&& \hspace*{-4em}
{\cal M}^{\si_+,\si_-,\si_1,\si_2,\si_3,\si_4}_{\mathrm{NCd}}
(p_+,p_-,k_1,k_2,k_3,k_4) 
\nn\\* &=& \phantom{{}+{}}
{\cal M}^{\si_+,\si_-,\si_1,\si_2,\si_3,\si_4}_{\mathrm{NCa}}
(p_+,p_-,k_1,k_2,k_3,k_4) 
\nn\\ && {}
- {\cal M}^{\si_+,\si_-,\si_3,\si_2,\si_1,\si_4}_{\mathrm{NCa}}
(p_+,p_-,k_3,k_2,k_1,k_4)
\nn\\ && {}
- {\cal M}^{-\si_1,\si_-,-\si_+,\si_2,\si_3,\si_4}_{\mathrm{NCa}}
(-k_1,p_-,-p_+,k_2,k_3,k_4)
\nn\\ && {}
- {\cal M}^{-\si_3,\si_-,\si_1,\si_2,-\si_+,\si_4}_{\mathrm{NCa}}
(-k_3,p_-,k_1,k_2,-p_+,k_4) 
\nn\\ && {}
+ {\cal M}^{-\si_1,\si_-,\si_3,\si_2,-\si_+,\si_4}_{\mathrm{NCa}}
(-k_1,p_-,k_3,k_2,-p_+,k_4) 
\nn\\ && {}
+ {\cal M}^{-\si_3,\si_-,-\si_+,\si_2,\si_1,\si_4}_{\mathrm{NCa}}
(-k_3,p_-,-p_+,k_2,k_1,k_4).
\eeqar
Finally, the helicity amplitudes for reactions of mixed CC/NC type read
\beqar
&& \hspace*{-4em}
{\cal M}^{\si_+,\si_-,\si_1,\si_2,\si_3,\si_4}_{\mathrm{CC/NCa}}
(p_+,p_-,k_1,k_2,k_3,k_4) 
\nn\\ &=& \phantom{{}+{}}
{\cal M}^{\si_+,\si_-,\si_1,\si_2,\si_3,\si_4}_{\mathrm{NCa}}
(p_+,p_-,k_1,k_2,k_3,k_4) 
\nn\\ && {}
- {\cal M}^{\si_+,\si_-,\si_1,\si_4,\si_3,\si_2}_{\mathrm{CCa}}
(p_+,p_-,k_1,k_4,k_3,k_2),
\\[1em]
&& \hspace*{-4em}
{\cal M}^{\si_+,\si_-,\si_1,\si_2,\si_3,\si_4}_{\mathrm{CC/NCb}}
(p_+,p_-,k_1,k_2,k_3,k_4) 
\nn\\ &=& \phantom{{}+{}}
{\cal M}^{\si_+,\si_-,\si_1,\si_2,\si_3,\si_4}_{\mathrm{NCa}}
(p_+,p_-,k_1,k_2,k_3,k_4) 
\nn\\ && {}
- {\cal M}^{-\si_3,-\si_4,\si_1,-\si_-,-\si_+,\si_2}_{\mathrm{CCa}}
(-k_3,-k_4,k_1,-p_-,-p_+,k_2),
\\[1em]
&& \hspace*{-4em}
{\cal M}^{\si_+,\si_-,\si_1,\si_2,\si_3,\si_4}_{\mathrm{CC/NCc}}
(p_+,p_-,k_1,k_2,k_3,k_4) 
\nn\\ &=& \phantom{{}+{}}
{\cal M}^{\si_+,\si_-,\si_1,\si_2,\si_3,\si_4}_{\mathrm{NCa}}
(p_+,p_-,k_1,k_2,k_3,k_4) 
\nn\\ && {}
- {\cal M}^{\si_+,\si_-,\si_3,\si_2,\si_1,\si_4}_{\mathrm{NCa}}
(p_+,p_-,k_3,k_2,k_1,k_4) 
\nn\\ && {}
- {\cal M}^{-\si_1,-\si_2,-\si_+,\si_4,\si_3,-\si_-}_{\mathrm{CCa}}
(-k_1,-k_2,k_3,-p_-,-p_+,k_4) 
\nn\\ && {}
+ {\cal M}^{-\si_1,-\si_4,-\si_+,\si_2,\si_3,-\si_-}_{\mathrm{CCa}}
(-k_1,-k_4,k_3,-p_-,-p_+,k_2) 
\nn\\ && {}
+ {\cal M}^{-\si_3,-\si_2,-\si_+,\si_4,\si_1,-\si_-}_{\mathrm{CCa}}
(-k_3,-k_2,k_1,-p_-,-p_+,k_4) 
\nn\\ && {}
- {\cal M}^{-\si_3,-\si_4,-\si_+,\si_2,\si_1,-\si_-}_{\mathrm{CCa}}
(-k_3,-k_4,k_1,-p_-,-p_+,k_2),
\\[1em]
&& \hspace*{-4em}
{\cal M}^{\si_+,\si_-,\si_1,\si_2,\si_3,\si_4}_{\mathrm{CC/NCd}}
(p_+,p_-,k_1,k_2,k_3,k_4) 
\nn\\ &=& \phantom{{}+{}}
{\cal M}^{\si_+,\si_-,\si_1,\si_2,\si_3,\si_4}_{\mathrm{NCa}}
(p_+,p_-,k_1,k_2,k_3,k_4) 
\nn\\ && {}
- {\cal M}^{-\si_3,\si_-,\si_1,\si_2,-\si_+,\si_4}_{\mathrm{NCa}}
(-k_3,p_-,k_1,k_2,-p_+,k_4) 
\nn\\ && {}
- {\cal M}^{\si_+,\si_-,\si_1,\si_4,\si_3,\si_2}_{\mathrm{CCa}}
(p_+,p_-,k_1,k_4,k_3,k_2) 
\nn\\ && {}
+ {\cal M}^{\si_+,-\si_4,\si_1,-\si_-,\si_3,\si_2}_{\mathrm{CCa}}
(p_+,-k_4,k_1,-p_-,k_3,k_2) 
\nn\\ && {}
+ {\cal M}^{-\si_3,\si_-,\si_1,\si_4,-\si_+,\si_2}_{\mathrm{CCa}}
(-k_3,p_-,k_1,k_4,-p_+,k_2) 
\nn\\ && {}
- {\cal M}^{-\si_3,-\si_4,\si_1,-\si_-,-\si_+,\si_2}_{\mathrm{CCa}}
(-k_3,-k_4,k_1,-p_-,-p_+,k_2).
\eeqar
The relative signs between contributions of the basic subamplitudes 
${\cal M}_{\mathrm{CCa}}$ and ${\cal M}_{\mathrm{NCa}}$ to the full matrix
elements account for the sign changes resulting from interchanging
external fermion lines.

For the CC reactions, the amplitudes $\M_{\mathrm{CCa}}$ are the
smallest gauge-invariant subset of diagrams \cite{Bo99}. In the
case of NC reactions, the amplitudes $\M_{\mathrm{NCa}}$ are composed of
three separately gauge-invariant subamplitudes consisting of the first
two lines, the two lines in the middle, and the last two lines of
\refeq{eq:ee4f_nca}.

\subsubsection{Hadronic final states}
\label{se:hadfinstat}

\begin{sloppypar}
  Next we inspect purely hadronic final states, i.e.\ the cases where
  all final-state fermions $f_i$ are quarks. This concerns only the
  channels CC(a), NC(a), NC(b), and CC/NC(a) given in
  \refse{se:pclass}. The colour structure of the quarks leads to two
  kinds of modifications. Firstly, the summation of the squared
  amplitudes over the colour degrees of freedom can become
  non-trivial, and secondly, the possibility of virtual-gluon exchange
  between the quarks has to be taken into account.  More precisely,
  there are diagrams of type (a) in \reffi{ee4fdiags} in which one of
  the gauge bosons $V_{1,2}$ is a gluon. The other gauge boson of
  $V_{1,2}$ can only be a photon or Z~boson, since this boson has to
  couple to the incoming $\Pep\Pem$ pair. Consequently, there is an
  impact of gluon-exchange diagrams only for the channels NC(a),
  NC(b), and CC/NC(a), but not for CC(a).  This can be easily seen by
  inspecting the generic diagrams in \reffi{ee4fdiags}: the presence
  of a gluon exchange requires two quark--antiquark pairs $q\bar q$ in
  the final state.
\end{sloppypar}

We first inspect the colour structure of the purely electroweak
diagrams. Since the colour structure of each diagram contributing to the
basic channels CC(a) and NC(a) is the same, the corresponding amplitudes
factorize into a simple colour part and the ``colour-singlet
amplitudes'' ${\cal M}_{\mathrm{CCa}}$ and ${\cal M}_{\mathrm{NCa}}$,
given in \refeq{eq:ee4f_cca} and \refeq{eq:ee4f_nca}, respectively.
The amplitudes for NC(b) and CC/NC(a) are composed from the ones of CC(a)
and NC(a) in a way that is analogous to the singlet case, but now the
colour indices $c_i$ of the quarks $f_i$ have to be taken into account.
Indicating the electroweak amplitudes for fully hadronic final states 
by ``$\mathrm{had,ew}$'', and writing colour indices explicitly, we get
\beqar
\lefteqn{
{\cal M}^{\si_+,\si_-,\si_1,\si_2,\si_3,\si_4}
_{{\mathrm{CCa,had,ew,}}c_1,c_2,c_3,c_4}
(p_+,p_-,k_1,k_2,k_3,k_4) } \hspace*{7em} &&
\nn\\
&=& \phantom{{}+{}}
{\cal M}^{\si_+,\si_-,\si_1,\si_2,\si_3,\si_4}_{\mathrm{CCa}}
(p_+,p_-,k_1,k_2,k_3,k_4) \, \delta_{c_1 c_2} \delta_{c_3 c_4},
\\[1em]
\lefteqn{
{\cal M}^{\si_+,\si_-,\si_1,\si_2,\si_3,\si_4}
_{{\mathrm{NCa,had,ew,}}c_1,c_2,c_3,c_4}
(p_+,p_-,k_1,k_2,k_3,k_4) } \hspace*{7em} &&
\nn\\
&=& \phantom{{}+{}}
{\cal M}^{\si_+,\si_-,\si_1,\si_2,\si_3,\si_4}_{\mathrm{NCa}}
(p_+,p_-,k_1,k_2,k_3,k_4) \, \delta_{c_1 c_2} \delta_{c_3 c_4},
\\[1em]
\lefteqn{
{\cal M}^{\si_+,\si_-,\si_1,\si_2,\si_3,\si_4}
_{{\mathrm{NCb,had,ew,}}c_1,c_2,c_3,c_4}
(p_+,p_-,k_1,k_2,k_3,k_4) } \hspace*{7em} &&
\nn\\* 
&=& \phantom{{}+{}}
{\cal M}^{\si_+,\si_-,\si_1,\si_2,\si_3,\si_4}_{\mathrm{NCa}}
(p_+,p_-,k_1,k_2,k_3,k_4) \, \delta_{c_1 c_2} \delta_{c_3 c_4}
\nn\\ && {}
- {\cal M}^{\si_+,\si_-,\si_3,\si_2,\si_1,\si_4}_{\mathrm{NCa}}
(p_+,p_-,k_3,k_2,k_1,k_4) \, \delta_{c_3 c_2} \delta_{c_1 c_4},
\label{eq:ee4f_ncbhadew}
\\[1em]
\lefteqn{
{\cal M}^{\si_+,\si_-,\si_1,\si_2,\si_3,\si_4}
_{{\mathrm{CC/NCa,had,ew,}}c_1,c_2,c_3,c_4}
(p_+,p_-,k_1,k_2,k_3,k_4) } \hspace*{7em} &&
\nn\\*
&=& \phantom{{}+{}}
{\cal M}^{\si_+,\si_-,\si_1,\si_2,\si_3,\si_4}_{\mathrm{NCa}}
(p_+,p_-,k_1,k_2,k_3,k_4) \, \delta_{c_1 c_2} \delta_{c_3 c_4}
\nn\\ && {}
- {\cal M}^{\si_+,\si_-,\si_1,\si_4,\si_3,\si_2}_{\mathrm{CCa}}
(p_+,p_-,k_1,k_4,k_3,k_2) \, \delta_{c_1 c_4} \delta_{c_3 c_2}.
\label{eq:ee4f_ccncahadew}
\eeqar

In the calculation of the gluon-exchange diagrams we can also make use
of the ``colour-singlet'' result \refeq{genfun4fNC} for the generic diagram 
(a) of \reffi{ee4fdiags}, after splitting off the colour structure
appropriately. Since each of these diagrams
involves exactly one internal gluon,
exchanged by the two quark lines, the corresponding matrix elements can
be deduced in a simple way from the diagrams in which the gluon is
replaced by a photon. The 
gluon-exchange contributions to the channels NC(b)
and CC/NC(a) can again be composed from the ones for NC(a). Making use of
the auxiliary function
\beqar
\lefteqn{
{\cal M}^{\si_+,\si_-,\si_1,\si_2,\si_3,\si_4}_{\mathrm{g}}
(p_+,p_-,k_1,k_2,k_3,k_4) 
} \hspace*{3em} &&
\nn\\* 
& = \disp\frac{g_{\mathrm{s}}^2}{Q_1 Q_3 e^2} \,
\sum_{V=\gamma,Z} \Bigl[ & \phantom{{}+{}}
{\cal M}^{-\si_1,-\si_2,\si_+,\si_-,-\si_3,-\si_4}_{V\gamma}
(-k_1,-k_2,p_+,p_-,-k_3,-k_4)
\nn\\ && {}
+ {\cal M}^{-\si_3,-\si_4,\si_+,\si_-,-\si_1,-\si_2}_{V\gamma}
(-k_3,-k_4,p_+,p_-,-k_1,-k_2)
\nn\\ && {}
+ {\cal M}^{-\si_1,-\si_2,-\si_3,-\si_4,\si_+,\si_-}_{\gamma V}
(-k_1,-k_2,-k_3,-k_4,p_+,p_-)
\nn\\ && {}
+ {\cal M}^{-\si_3,-\si_4,-\si_1,-\si_2,\si_+,\si_-}_{\gamma V}
(-k_3,-k_4,-k_1,-k_2,p_+,p_-) \Bigl],
\hspace*{2em}
\eeqar
where $g_{\mathrm{s}}=\sqrt{4\pi\alpha_{\mathrm{s}}}$ is the strong
gauge coupling, the matrix elements involving gluon exchange
explicitly read
\beqar
\lefteqn{
{\cal M}^{\si_+,\si_-,\si_1,\si_2,\si_3,\si_4}
_{{\mathrm{NCa,had,gluon,}}c_1,c_2,c_3,c_4}
(p_+,p_-,k_1,k_2,k_3,k_4) } \hspace*{7em} &&
\nn\\* 
&=& \phantom{{}+{}}
{\cal M}^{\si_+,\si_-,\si_1,\si_2,\si_3,\si_4}_{\mathrm{g}}
(p_+,p_-,k_1,k_2,k_3,k_4) 
\; \text\frac{1}{4} \, \lambda^a_{c_1 c_2} \lambda^a_{c_3 c_4},
\\[1em]
\lefteqn{
{\cal M}^{\si_+,\si_-,\si_1,\si_2,\si_3,\si_4}
_{{\mathrm{NCb,had,gluon,}}c_1,c_2,c_3,c_4}
(p_+,p_-,k_1,k_2,k_3,k_4) } \hspace*{7em} &&
\nn\\* 
&=& \phantom{{}+{}}
{\cal M}^{\si_+,\si_-,\si_1,\si_2,\si_3,\si_4}_{\mathrm{g}}
(p_+,p_-,k_1,k_2,k_3,k_4) 
\; \text\frac{1}{4} \, \lambda^a_{c_1 c_2} \lambda^a_{c_3 c_4}
\nn\\ && {}
- {\cal M}^{\si_+,\si_-,\si_3,\si_2,\si_1,\si_4}_{\mathrm{g}}
(p_+,p_-,k_3,k_2,k_1,k_4)
\; \text\frac{1}{4} \, \lambda^a_{c_3 c_2} \lambda^a_{c_1 c_4},
\label{eq:ee4f_ncbhadqcd}
\\[1em]
\lefteqn{
{\cal M}^{\si_+,\si_-,\si_1,\si_2,\si_3,\si_4}
_{{\mathrm{CC/NCa,had,gluon,}}c_1,c_2,c_3,c_4}
(p_+,p_-,k_1,k_2,k_3,k_4) } \hspace*{7em} &&
\nn\\*
&=& \phantom{{}+{}}
{\cal M}^{\si_+,\si_-,\si_1,\si_2,\si_3,\si_4}_{\mathrm{g}}
(p_+,p_-,k_1,k_2,k_3,k_4) 
\; \text\frac{1}{4} \, \lambda^a_{c_1 c_2} \lambda^a_{c_3 c_4}.
\eeqar
The colour structure is easily evaluated by making use of the
completeness relation
$\lambda^a_{ij} \lambda^a_{kl} = -\text\frac{2}{3}\delta_{ij}\delta_{kl}
+2\delta_{il}\delta_{jk}$ for the Gell-Mann matrices $\lambda^a_{ij}$.

\begin{sloppypar}
The complete matrix elements for the fully hadronic channels result
from the sum of the 
purely electroweak and the gluon-exchange  contributions,
\beq
{\cal M}^{\si_+,\si_-,\si_1,\si_2,\si_3,\si_4}
_{{\ldots\mathrm{,had,}}c_1,c_2,c_3,c_4}
= {\cal M}^{\si_+,\si_-,\si_1,\si_2,\si_3,\si_4}
_{{\ldots\mathrm{,had,ew,}}c_1,c_2,c_3,c_4}+
{\cal M}^{\si_+,\si_-,\si_1,\si_2,\si_3,\si_4}
_{{\ldots\mathrm{,had,gluon,}}c_1,c_2,c_3,c_4}.
\eeq
The gluon-exchange contributions are separately gauge-invariant.

For clarity, we explicitly write down the colour-summed squared matrix
elements for the fully hadronic channels. Abbreviating 
${\cal M}_{\dots}
^{\si_+,\si_-,\si_a,\si_b,\si_c,\si_d}(p_+,p_-,k_a,k_b,k_c,k_d)$
by ${\cal M}_{\dots}(a,b,c,d)$ we obtain
\end{sloppypar}
\beqar
\lefteqn{
\sum_{\mathrm{colour}}|{\cal M}_{\mathrm{CCa,had}}(1,2,3,4)|^2 
= 9 |{\cal M}_{\mathrm{CCa}}(1,2,3,4)|^2,
} \hspace*{0em} &&
\\[1em]
\lefteqn{
\sum_{\mathrm{colour}}|{\cal M}_{\mathrm{NCa,had}}(1,2,3,4)|^2 
= 9 |{\cal M}_{\mathrm{NCa}}(1,2,3,4)|^2
+ 2 |{\cal M}_{\mathrm{g}}(1,2,3,4)|^2,
} \hspace*{0em} &&
\\[1em]
\lefteqn{
\sum_{\mathrm{colour}}|{\cal M}_{\mathrm{NCb,had}}(1,2,3,4)|^2 
} \hspace*{0em} &&
\nn\\*
&=& 9 |{\cal M}_{\mathrm{NCa}}(1,2,3,4)|^2
+ 9 |{\cal M}_{\mathrm{NCa}}(3,2,1,4)|^2
- 6\Re\big\{ {\cal M}_{\mathrm{NCa}}(1,2,3,4)
{\cal M}^*_{\mathrm{NCa}}(3,2,1,4) \big\}
\nn\\ && {}
+ 2 |{\cal M}_{\mathrm{g}}(1,2,3,4)|^2
+ 2 |{\cal M}_{\mathrm{g}}(3,2,1,4)|^2
+ \text\frac{4}{3}\Re\big\{ {\cal M}_{\mathrm{g}}(1,2,3,4)
{\cal M}^*_{\mathrm{g}}(3,2,1,4) \big\}
\nn\\ && {}
- 8\Re\big\{ {\cal M}_{\mathrm{NCa}}(1,2,3,4)
{\cal M}^*_{\mathrm{g}}(3,2,1,4) \big\}
- 8\Re\big\{ {\cal M}_{\mathrm{NCa}}(3,2,1,4)
{\cal M}^*_{\mathrm{g}}(1,2,3,4) \big\},
\nn\\*
\\[1em]
\lefteqn{
\sum_{\mathrm{colour}}|{\cal M}_{\mathrm{CC/NCa,had}}(1,2,3,4)|^2 
} \hspace*{0em} &&
\nn\\
&=& 9 |{\cal M}_{\mathrm{NCa}}(1,2,3,4)|^2
+ 9 |{\cal M}_{\mathrm{CCa}}(1,4,3,2)|^2
- 6\Re\big\{ {\cal M}_{\mathrm{NCa}}(1,2,3,4)
{\cal M}^*_{\mathrm{CCa}}(1,4,3,2) \big\}
\nn\\ && {}
+ 2 |{\cal M}_{\mathrm{g}}(1,2,3,4)|^2
- 8\Re\big\{ {\cal M}_{\mathrm{CCa}}(1,4,3,2)
{\cal M}^*_{\mathrm{g}}(1,2,3,4) \big\}.
\eeqar 
Owing to the colour structure of the diagrams, a non-zero interference
between purely electroweak and gluon-exchange contributions is only
possible if the four final-state fermions can be combined into one
single closed fermion line in the squared diagram. This implies that
fermion pairs must couple to different resonances in the
electroweak and the gluon-exchange diagrams,
leading to a global suppression of such interference effects in the
phase-space integration (see \refse{se:QCD}).

\subsection{Generic functions and amplitudes for 
\boldmath{$\eeffffg$} }
\label{se:genfunction}

The generic functions for $\eeffffg$ can be constructed in a similar
way. The idea is to combine the contributions of all those graphs to
one generic function that reduce to the same graph after removing the
radiated photon.  These combined contributions to $\eeffffg$ are
classified in the same way as the diagrams for the corresponding
process $\Pep\Pem\to 4f$, i.e.\ the graphs of \reffi{ee4fdiags} also
represent the generic functions for $\eeffffg$. Finally, all
amplitudes for $\eeffffg$ can again be constructed from only two
generic functions.  Note that the number of individual Feynman
diagrams ranges between 14 and 1008 for the various processes.  We
note that the generic functions can in fact be used to construct the
amplitudes for all processes involving exactly six external fermions
and one external photon, such as $\Pem\Pem\to4f\ga$ and $\Pe\ga\to5f$.

As a virtue of this approach, the so-defined generic functions fulfill
the QED Ward identity for the external photon, i.e.\ replacing the
photon polarization vector by the photon momentum yields zero for each
generic function. This is simply a consequence of electromagnetic
charge conservation. Consequently, in the actual calculation in the
WvdW formalism the gauge spinor of the photon drops out in each
contribution separately.

Assuming the external fermions as incoming and the photon as outgoing, 
the generic functions read
\beqar
&& \hspace*{-3em}
{\cal M}^{\si_a,\si_b,\si_c,\si_d,\si_e,\si_f,\lambda}_{V_1 V_2}
(Q_a,Q_b,Q_c,Q_d,Q_e,Q_f,p_a,p_b,p_c,p_d,p_e,p_f,k) 
\nn\\*
\hspace*{2em}
&=& -4\sqrt{2}e^5 \de_{\si_a,-\si_b} \de_{\si_c,-\si_d} \de_{\si_e,-\si_f} \, 
g^{\si_b}_{V_1\bar f_a f_g} g^{\si_b}_{V_2\bar f_g f_b}
g^{\si_d}_{V_1\bar f_c f_d} g^{\si_f}_{V_2\bar f_e f_f}
\nn\\ && {} \times
A^{\si_a,\si_c,\si_e,\lambda}_2
(Q_a,Q_b,Q_c,Q_d,Q_e,Q_f,p_a,p_b,p_c,p_d,p_e,p_f,k),
\hspace*{3em}
\\[1em] 
&& \hspace*{-3em}
{\cal M}^{\si_a,\si_b,\si_c,\si_d,\si_e,\si_f,\lambda}_{VWW}
(Q_a,Q_b,Q_c,Q_d,Q_e,Q_f,p_a,p_b,p_c,p_d,p_e,p_f,k) 
\nn\\*
&=& -4\sqrt{2}e^5 \de_{\si_a,-\si_b} \de_{\si_c,+} \de_{\si_d,-} 
\de_{\si_e,+} \de_{\si_f,-} \, 
(Q_c-Q_d)g_{VWW} g^{\si_b}_{V\bar f_a f_b} 
g^{-}_{W\bar f_c f_d} g^{-}_{W\bar f_e f_f}
\nn\\ && {} \times
A^{\si_a,\lambda}_3
(Q_a,Q_b,Q_c,Q_d,Q_e,Q_f,p_a,p_b,p_c,p_d,p_e,p_f,k),
\label{eq:Mgenee4fA}
\eeqar
with the auxiliary functions
\beqar
&& \hspace*{-3em}
A^{{+}{+}{+}{+}}_2(Q_a,Q_b,Q_c,Q_d,Q_e,Q_f,p_a,p_b,p_c,p_d,p_e,p_f,k) 
= -\spac \biggl\{
\nn\\
\hspace*{2em} && 
P_{V_1}(p_c+p_d) P_{V_2}(p_e+p_f) 
\nn\\ 
&& \quad
{} \times \biggl[ \frac{\cspbf}{\spak} \biggl(
\frac{Q_c-Q_d}{(p_b+p_e+p_f)^2} \frac{\spac}{\spck}
(\cspbd \spbe+\cspdf \spef)
\nn\\ 
&& \quad
\hphantom{{} \times \biggl[ \frac{\cspbf}{\spak} \biggl(} {}
+\frac{Q_f-Q_e}{(p_a+p_c+p_d)^2} \frac{\spae}{\spek}
(\cspad \spae+\cspcd \spce) \biggr)
\nn\\ 
&& \quad \hphantom{\times \biggl[}
+\frac{Q_b (\cspad\spae+\cspcd\spce) (\cspbf\spab-\cspfk\spak) }
{(p_a+p_c+p_d)^2\spak\spbk} 
\nn\\ 
&& \quad \hphantom{\times \biggl[}
+\frac{(Q_a+Q_c-Q_d) \cspbf\cspcd\spac (\cspbk\spbe-\cspfk\spef) }
{(p_a+p_c+p_d)^2 (p_b+p_e+p_f)^2 \spak} \biggr]
\nn\\ 
&& {}
-\frac{Q_d-(Q_c-Q_d) 2(p_d\cdot k) P_{V_1}(p_c+p_d)}{(p_b+p_e+p_f)^2}
P_{V_1}(p_c+p_d-k) P_{V_2}(p_e+p_f) \cspbf
\nn\\ 
&& {} \hphantom{{}+{}} \;
\times \frac{ \spcd (\cspbd \spbe+\cspdf \spef)
+\spck (\cspbk \spbe-\cspfk \spef) } {\spck\spdk} 
\nn\\ 
&& {}
+\frac{Q_f-(Q_e-Q_f) 2(p_f\cdot k) P_{V_2}(p_e+p_f)}{(p_a+p_c+p_d)^2}
P_{V_1}(p_c+p_d) P_{V_2}(p_e+p_f-k)
\nn\\ 
&& {} \hphantom{{}+{}} \;
\times \frac{ (\cspbf \spef+\cspbk \spek) (\cspad \spae+\cspcd \spce)}
{\spek\spfk} 
\,\biggr\},
\nn\\
&& \hspace*{-3em}
A^{{+}{+}{-}{+}}_2(Q_a,Q_b,Q_c,Q_d,Q_e,Q_f,p_a,p_b,p_c,p_d,p_e,p_f,k)
\nn\\*
&=& A^{{+}{+}{+}{+}}_2(Q_a,Q_b,Q_c,Q_d,-Q_f,-Q_e,p_a,p_b,p_c,p_d,p_f,p_e,k),
\nn\\
&& \hspace*{-3em}
A^{{+}{-}{+}{+}}_2(Q_a,Q_b,Q_c,Q_d,Q_e,Q_f,p_a,p_b,p_c,p_d,p_e,p_f,k)
\nn\\*
&=& A^{{+}{+}{+}{+}}_2(Q_a,Q_b,-Q_d,-Q_c,Q_e,Q_f,p_a,p_b,p_d,p_c,p_e,p_f,k),
\nn\\
&& \hspace*{-3em}
A^{{+}{-}{-}{+}}_2(Q_a,Q_b,Q_c,Q_d,Q_e,Q_f,p_a,p_b,p_c,p_d,p_e,p_f,k)
\nn\\*
&=& A^{{+}{+}{+}{+}}_2(Q_a,Q_b,-Q_d,-Q_c,-Q_f,-Q_e,p_a,p_b,p_d,p_c,p_f,p_e,k),
\nn\\
&& \hspace*{-3em}
A^{{-}{+}{+}{+}}_2(Q_a,Q_b,Q_c,Q_d,Q_e,Q_f,p_a,p_b,p_c,p_d,p_e,p_f,k)
\nn\\*
&=& A^{{+}{+}{+}{+}}_2(Q_b,Q_a,-Q_e,-Q_f,-Q_c,-Q_d,p_b,p_a,p_e,p_f,p_c,p_d,k),
\nn\\
&& \hspace*{-3em}
A^{{-}{+}{-}{+}}_2(Q_a,Q_b,Q_c,Q_d,Q_e,Q_f,p_a,p_b,p_c,p_d,p_e,p_f,k)
\nn\\*
&=& A^{{+}{+}{+}{+}}_2(Q_b,Q_a,Q_f,Q_e,-Q_c,-Q_d,p_b,p_a,p_f,p_e,p_c,p_d,k),
\nn\\
&& \hspace*{-3em}
A^{{-}{-}{+}{+}}_2(Q_a,Q_b,Q_c,Q_d,Q_e,Q_f,p_a,p_b,p_c,p_d,p_e,p_f,k)
\nn\\*
&=& A^{{+}{+}{+}{+}}_2(Q_b,Q_a,-Q_e,-Q_f,Q_d,Q_c,p_b,p_a,p_e,p_f,p_d,p_c,k),
\nn\\
&& \hspace*{-3em}
A^{{-}{-}{-}{+}}_2(Q_a,Q_b,Q_c,Q_d,Q_e,Q_f,p_a,p_b,p_c,p_d,p_e,p_f,k)
\nn\\*
&=& A^{{+}{+}{+}{+}}_2(Q_b,Q_a,Q_f,Q_e,Q_d,Q_c,p_b,p_a,p_f,p_e,p_d,p_c,k),
\nn\\
&& \hspace*{-3em}
A^{\si_a,\si_c,\si_d,{-}}_2
(Q_a,Q_b,Q_c,Q_d,Q_e,Q_f,p_a,p_b,p_c,p_d,p_e,p_f,k)
\\*
&=& \Big( A^{-\si_a,-\si_c,-\si_d,{+}}_2
(Q_a,Q_b,Q_c,Q_d,Q_e,Q_f,p_a,p_b,p_c,p_d,p_e,p_f,k) \Bigr)^*
\Big|_{P_{V_{1,2}}(p)\to P_{V_{1,2}}^*(p)},
\nn
\label{eq:A2}
\eeqar
and
\beqar
&& \hspace*{-3em}
A^{{+}{+}}_3(Q_a,Q_b,Q_c,Q_d,Q_e,Q_f,p_a,p_b,p_c,p_d,p_e,p_f,k) 
\nn\\*
\hspace*{2em} 
&=& P_V(p_a+p_b) P_W(p_c+p_d) P_W(p_e+p_f) 
\frac{(Q_c-Q_d)\spce}{\spck\spek} 
\nn\\
&& \quad
{} \times ( \hphantom{{}+{}} \cspbd\cspbf\spab\spce +\cspbd\cspdf\spae\spcd 
\nn\\
&& \quad \hphantom{{} \times (}
{} + \cspbf\cspdf\spac\spef )
\nn\\
&& {} +P_V(p_a+p_b-k) P_W(p_c+p_d) P_W(p_e+p_f) \frac{Q_b}{\spak\spbk} 
\nn\\
&& \quad
{} \times \{ \hphantom{{}+{}} \cspdf [ \hphantom{{}+{}}
\spae\spcd (\cspbd\spab-\cspdk\spak)
\nn\\
&& \quad \hphantom{{} \times \{ {}+ \cspdf [}
{} +\spac\spef (\cspbf\spab-\cspfk\spak) ]
\nn\\
&& \quad \hphantom{{} \times \{ }
{} + \spce (\cspbd\spab-\cspdk\spak) (\cspbf\spab-\cspfk\spak) \}
\nn\\
&& {} +P_V(p_a+p_b) P_W(p_c+p_d-k) P_W(p_e+p_f) 
\frac{Q_d-(Q_c-Q_d)2(p_d\cdot k) P_W(p_c+p_d)}{\spck\spdk} 
\nn\\
&& \quad
{} \times \{ \hphantom{{}+{}} 
\cspbf [ \hphantom{{}+{}} \spac\spef (\cspdf\spcd-\cspfk\spck)
\nn\\
&& \quad \hphantom{{} \times \{ \hphantom{{}+{}} \cspbf [ }
{} +\spce\spab (\cspbd\spcd+\cspbk\spck) ]
\nn\\
&& \quad \hphantom{{} \times \{ }
{} + \spae (\cspdf\spcd-\cspfk\spck) (\cspbd\spcd+\cspbk\spck) \}
\nn\\
&& {} +P_V(p_a+p_b) P_W(p_c+p_d) P_W(p_e+p_f-k) 
\frac{Q_f+(Q_f-Q_e)2(p_f\cdot k) P_W(p_e+p_f)}{\spek\spfk} 
\nn\\
&& \quad
{} \times \{ \hphantom{{}+{}} 
\cspbd [ \hphantom{{}+{}} \spce\spab (\cspbf\spef+\cspbk\spek)
\nn\\
&& \quad \phantom{ {} \times \{ \hphantom{{}+{}} \cspbd [ }
{} + \spae\spcd (\cspdf\spef+\cspdk\spek) ]
\nn\\
&& \quad \phantom{ {} \times \{ }
{}+\spac (\cspbf\spef+\cspbk\spek) (\cspdf\spef+\cspdk\spek) \},
\nn\\
&& \hspace*{-3em}
A^{{-}{+}}_3(Q_a,Q_b,Q_c,Q_d,Q_e,Q_f,p_a,p_b,p_c,p_d,p_e,p_f,k) 
\nn\\*
&=& A^{{+}{+}}_3(-Q_b,-Q_a,Q_c,Q_d,Q_e,Q_f,p_b,p_a,p_c,p_d,p_e,p_f,k),
\nn\\
&& \hspace*{-3em}
A^{\si_a,{-}}_3(Q_a,Q_b,Q_c,Q_d,Q_e,Q_f,p_a,p_b,p_c,p_d,p_e,p_f,k) 
\\*
&=& \Bigl( 
A^{-\si_a,{+}}_3(Q_a,Q_b,-Q_d,-Q_c,-Q_f,-Q_e,p_a,p_b,p_d,p_c,p_f,p_e,k) 
\Bigr)^* \Big|_{P_{V,W}(p)\to P_{V,W}^*(p)}.
\nn
\label{eq:A3}
\eeqar
The replacements $P_V\to P_V^*$ after the complex conjugation
in the last lines of \refeq{eq:A2} and \refeq{eq:A3} ensure that the
vector-boson propagators remain unaffected.
Note that the vector-boson masses do not enter explicitly in the above
results, but only via $P_V$. 
In gauges such as the 't~Hooft--Feynman or the unitary gauge this feature 
is obtained only after combining different Feynman graphs for $\eeffffg$; 
in the
non-linear gauge \refeq{eq:nlgauge} this is the case diagram by diagram.

The helicity amplitudes for $\eeffffg$ follow from the
generic functions ${\cal M}_{V_1 V_2}$ and ${\cal M}_{VWW}$ of 
\refeq{eq:Mgenee4fA} in exactly the same way as described in
\refse{se:ee4f} 
for $\Pep\Pem\to 4f$.
This holds also for the gluon-exchange matrix elements and for the
colour factors.
Moreover, the classification of gauge-invariant sets of diagrams
for $\eeffff$ immediately yields such sets for $\eeffffg$, if the additional
photon is attached to all graphs of a set in all possible ways.

We have checked analytically that the electromagnetic Ward identity
for the external photon is fulfilled for each generic contribution
separately. In addition, we have numerically compared the amplitudes
for all processes with amplitudes generated by {\sl Madgraph} \cite{St94}
 for zero width of the vector bosons and found complete agreement.
We could not compare our results with {\sl Madgraph} for finite width,
because {\sl Madgraph} uses the unitary gauge for massive vector-boson
propagators and the `t~Hooft--Feynman gauge for the photon propagators,
while we are using the non-linear gauge \refeq{eq:nlgauge}. Therefore,
the matrix elements differ after introduction of finite vector-boson widths.
While the calculation with {\sl Madgraph} is  fully automized,
in our calculation we have full control over the matrix
element and can, in particular, investigate various implementations of
the finite width.

A comparison of our results with those of \citeres{Ae91,Ae91a}, which include
only the matrix elements that involve two resonant \PW~bosons,
immediately reveals the virtues of our generic approach.

\subsection{Implementation of finite gauge-boson widths}
\label{se:finwidth}

We have implemented the finite widths of the $\PW$ and $\PZ$ bosons in
different ways:
\begin{itemize}
\item{\it fixed width} in all propagators: 
$P_V(p) = [p^2-\MV^2+\ri\MV\GV]^{-1}$,
\item{\it running width} in time-like propagators: 
$P_V(p) = [p^2-\MV^2+\ri p^2(\GV/\MV)\theta(p^2)]^{-1}$,
\item{\it complex-mass scheme:} complex gauge-boson masses everywhere,
  \ie $\sqrt{M_V^2-\ri M_V\Ga_V}$ instead of $M_V$ in the propagators
  and in the couplings. This results, in particular, in a constant
  width in all propagators,
\beq\label{Vprop}
  P_V(p) = [p^2-\MV^2+\ri\MV\Ga_V]^{-1},
\eeq
and in a complex weak mixing angle: 
\beq\label{complangle}
\cw^2=1-\sw^2=   \frac{\MW^2-\ri\MW\GW}{\MZ^2-\ri\MZ\GZ}.
\eeq
\end{itemize}
The virtues and drawbacks of the first two schemes
have been discussed in \citere{bhf2}. 
Both violate $\SU(2)$ gauge invariance, the running width
also $\U(1)$ gauge invariance.  
The complex-mass scheme obeys
all Ward identities%
\footnote{In the context of electromagnetic gauge invariance, the
introduction of a complex mass has been proposed in \citere{Lo91}}
and thus gives a consistent description of the
finite-width effects in any tree-level calculation.  While the 
complex-mass scheme 
works in general, it is particularly simple for $\eeffffg$ in the
non-linear gauge \refeq{eq:nlgauge}.  In this case, no couplings
involving explicit gauge-boson masses appear,
and it is sufficient to
introduce the finite gauge-boson widths in the propagators [\cf
\refeq{Vprop}] and to introduce the complex weak mixing angle
\refeq{complangle} in the couplings.  We note that a generalization of
this scheme to higher orders requires to introduce complex mass
counterterms in order to compensate for the complex masses in the
propagators \cite{St90}. We did not consider the
fermion-loop scheme \cite{bhf2,FLscheme}, which is also fully
consistent for lowest-order predictions, since it requires the
calculation of fermionic one-loop corrections to $\eeffffg$ 
which is beyond the scope of this work.

\section{The Monte Carlo programs}
\label{se:MC}

The cross section for $\Pep\Pem\to 4f(\gamma)$ is given by
\begin{eqnarray}\label{eq:crosssection}
\rd \sigma &=& \frac{(2 \pi)^{4-3 n}}{2 s}
\left[\prod\limits_{i=1}^n \rd^4 k_i \, 
\delta\left(k_i^2\right) \theta(k_i^0)\right]
\delta^{(4)} \left( p_+ +p_- -\sum_{i=1}^n k_i \right)  
\nonumber \\
&& {} 
\times |{\cal M}(p_+,p_-,k_1,\ldots ,k_n)|^2,
\end{eqnarray}
where 
$n=4,5$ is the number of outgoing particles. 
The helicity amplitudes ${\cal M}$ for $\Pep\Pem\to 4f(\gamma)$
have been calculated in \refses{se:ee4f} and \ref{se:genfunction}, 
respectively. 
The phase-space integration is performed
with the help of a Monte Carlo technique, since
the Monte Carlo method allows 
us to calculate a variety of observables
simultaneously and to 
easily implement cuts in order to
account for the experimental situation.

The helicity amplitudes in \refeq{eq:crosssection} 
exhibit a complicated peaking behaviour in different
regions of the integration domain. In order to obtain a numerically stable
result and to reduce the Monte Carlo integration error 
we use a multi-channel Monte Carlo method \cite{Be94,Multichannel}, which 
is briefly outlined in the following.   

Before turning to the multi-channel method, we consider the treatment
of a single channel. We choose a suitable set $\vec{\Phi}$ of $3n-4$
phase-space variables to describe a point in phase space, and
determine the corresponding physical region $V$ and the relation
$k_i(\vec{\Phi})$ between the phase-space variables $\vec{\Phi}$ and
the momenta $k_1, \ldots k_n$.  The phase-space integration of
\refeq{eq:crosssection} reads
\begin{eqnarray}
I_n &=& \int \rd \sigma=
\int_V \rd \vec{\Phi} \, \rho\Big(k_i(\vec{\Phi})\Big) \, 
f\Big(k_i(\vec{\Phi})\Big),
\\\nonumber
f\Big(k_i(\vec{\Phi})\Big)&=& \frac{(2 \pi)^{4-3 n}}{2 s}
\left|{\cal M}\left(p_+,p_-,k_1(\vec{\Phi}),\ldots ,
k_n(\vec{\Phi})\right)\right|^2,
\end{eqnarray}
where $\rho$ is the phase-space density. 
For the random generation of the events,
we further transform  
the integration variables $\vec{\Phi}$
to $3n-4$ new variables $\vec{r}=(r_i)$
with a hypercube as integration domain:
$\vec{\Phi}=\vec{h}(\vec{r}\,)$ with $0\le r_i \le 1$.
We obtain
\begin{equation}\label{eq:phint}
I_n = \int_V \rd \vec{\Phi} \, \rho\Big(k_i(\vec{\Phi})\Big) \, 
f\Big(k_i(\vec{\Phi})\Big) 
= \int_0^1 \rd \vec{r} \, 
\frac{f\Big(k_i(\vec{h}(\vec{r\,}))\Big)}
{g\Big(k_i(\vec{h}(\vec{r\,}))\Big)} \; ,
\end{equation}
where $g$ is the probability density of events generated in phase
space, defined by
\begin{equation} \label{eq:dens}
\frac{1}{g\Big(k_i(\vec{\Phi})\Big)} = \rho\Big(k_i(\vec{\Phi})\Big)
\left| \frac{\partial\vec{h}(\vec{r\,})}
{\partial\vec{r}} \right|_{\vec{r}=\vec{h}^{-1}(\vec{\Phi})} \; .
\end{equation} 
If $f$ varies strongly, the efficiency of the Monte Carlo method
can be considerably enhanced by choosing the
mapping of random numbers $\vec{r}$ into $\vec{\Phi}$ in such a way that 
the resulting density $g$ mimics the behaviour of $|f|$. 
For this {\it importance sampling}, the
choice of $\vec{\Phi}$ is guided by the peaking structure of $f$,
which is determined by the propagators in a characteristic Feynman diagram.

We choose the variables $\vec{\Phi}$ in such a way that the invariants
corresponding to the propagators are included.  Accordingly, we
decompose the $n$-particle final state into $2 \to 2$ scattering
processes with subsequent $1 \to 2$ decays.  The variables
$\vec{\Phi}$ consist of Lorentz invariants $s_i, t_i$, defined as the
squares of time- and space-like momenta, respectively, and of polar
and azimuthal angles $\theta_i,\phi_i$, defined in appropriate frames.
A detailed description of the parameterization of an $n$-particle
phase space in terms of invariants and angles can be found in
\citere{By73}.  The parameterization of the invariants $s_i,t_i$ in
$\vec{\Phi}=\vec{h}(\vec{r}\,)$ is chosen in such a way that the
propagator structure of the function $f$ is compensated by a similar
behaviour in the density $g$.  More precisely, if $f$ contains
Breit--Wigner resonances or distributions like $s_i^{-\nu}$, which are
relevant for massless propagators, appropriate parameterizations of
$s_i$ are given by:
\begin{itemize}
\item Breit--Wigner resonances:
\beqar\label{eq:maps1}
s_i &=& \MV^2+\MV\GV\tan[y_1+(y_2-y_1) r_i] 
\\ && \mbox{with } \;
y_{1,2}=\arctan\left(\frac{s_{\min,\max}-\MV^2}{\MV\GV}\right); \nn
\eeqar
\item propagators of massless particles:
\begin{eqnarray}\label{eq:maps2}
\nu\ne 1: \qquad s_i &=& \left[s_{\max}^{1-\nu} r_i
+s_{\min}^{1-\nu}(1-r_i)\right]^{1/(1-\nu)} ,
\nn\\[.5em]
\nu=1: \qquad
s_i &=&\exp\left[\ln(s_{\max}) r_i + \ln(s_{\min})(1-r_i)\right].
\end{eqnarray}
\end{itemize}
The parameter $\nu $ can be tuned to optimize the 
Monte Carlo integration and should be chosen 
$\gsim 1$.  The naive expectation $\nu =2$ is not necessarily the best
choice, because the propagator poles of the differential cross section
are partly cancelled in the collinear limit.  The remaining variables
in $\vec{\Phi}=\vec{h}(\vec{r}\,)$, i.e.\ those for which $f$ is
expected not to exhibit a peaking behaviour, are generated as follows:
\begin{eqnarray}\label{eq:nomaps}
s_i &=& s_{\max} r_i+s_{\min} (1-r_i), \qquad
\phi_i = 2 \pi r_i, \qquad \cos\theta_i =2 r_i-1.
\end{eqnarray}
The absolute values of the 
invariants $t_i$ are generated in the same way as $s_i$.
The resulting density $g$ of events in phase space 
is obtained as the product of the corresponding Jacobians,
as given in \refeq{eq:dens}.
In the appendix, we provide an explicit example for an event generation
with a specific choice of mappings $k_i(\vec{\Phi})$ and $\vec{h}(\vec{r}\,)$,
and for the calculation of the corresponding density $g$.

The differential cross sections of the processes
$\Pep\Pem\to 4f$ and especially $\eeffffg$ possess very
complex peaking structures so that the peaks in the integrand 
$f(\vec{\Phi})$ in \refeq{eq:phint} cannot be described properly 
by only one single density $g(\vec{\Phi})$.
The {\it multi-channel approach} \cite{Be94,Multichannel} 
suggests a solution to this problem. For each peaking structure we choose 
a suitable set $\vec{\Phi}_k$, and accordingly a 
mapping of random numbers $r_i$ into $\vec{\Phi}_k$:
$\vec{\Phi}_k=\vec{h}_k(\vec{r}\,)$ with $0 \le r_i \le 1$,
so that the resulting density
$g_k$ describes this particular peaking behaviour of $f$. 
All densities $g_k$ are combined 
into one density $g_{\mathrm {tot}}$
that is expected to smoothen the integrand over the whole phase-space 
integration region. The phase-space integral of \refeq{eq:phint} reads
\begin{eqnarray}
I_n &=& \sum_{k=1}^M \int_V \rd \vec{\Phi}_k \, 
\rho_k\Big(k_i(\vec{\Phi}_k)\Big)  \, 
g_k\Big(k_i(\vec{\Phi}_k)\Big) \,
\frac{f\Big(k_i(\vec{\Phi}_k)\Big)}{g_{\mathrm{tot}}\Big(k_i(\vec{\Phi}_k)\Big)}
=\sum_{k=1}^M \int_0^1 \rd \vec{r} \, 
\frac{f\Big(k_i(\vec{h}_k(\vec{r}\,))\Big)}
{g_{\mathrm {tot}}\Big(k_i(\vec{h}_k(\vec{r}\,))\Big)},
\hspace*{2em}
\eeqar
with
\beqar\label{eq:mdens}
g_{\mathrm {tot}}\Big(k_i(\vec{\Phi}_k)\Big)&=&
\sum_{l=1}^M g_l\Big(k_i(\vec{\Phi}_k)\Big)
,\qquad
\frac{1}{g_l\Big(k_i(\vec{\Phi}_k)\Big)}=\rho_l\Big(k_i(\vec{\Phi}_k)\Big) \, 
\left| \frac{\partial\vec{h}_l(\vec{r}\,)}
{\partial \vec{r}}\right|_{\vec{r}=\vec{h}_k^{-1}(\vec{\Phi}_k)}.
\end{eqnarray}
The different mappings $\vec{h}_k(\vec{r}\,)$ are called channels,
and $M$ is the number of all channels.

In order to reduce the Monte Carlo error further, we adopt the method
of weight optimization of \citere{Kl94} and 
introduce {\em a-priori weights} $\alpha_k, k=1,\dots, M$ ($\alpha_k \ge 0$
and $\sum_{k=1}^M \alpha_k=1$).
The channel $k$ that is used to generate the event is picked randomly
with probability $\alpha_k$, i.e.\ 
\begin{eqnarray}
I_n &=& \sum_{k=1}^M \alpha_k \int_V \rd \vec{\Phi}_k \, 
\rho_k\Big(k_i(\vec{\Phi}_k)\Big) 
g_k\Big(k_i(\vec{\Phi}_k)\Big) 
\, \frac{f\Big(k_i(\vec{\Phi}_k)\Big)}
{g_{\mathrm{tot}}\Big(k_i(\vec{\Phi}_k)\Big)}
\nonumber \\
&=& \int_0^1 \rd r_0 \, \sum_{k=1}^M 
\theta(r_0-\beta_{k-1})\theta(\beta_k-r_0)
\int_V \rd \vec{\Phi}_k \, \rho_k \Big(k_i(\vec{\Phi}_k)\Big) 
g_k\Big(k_i(\vec{\Phi}_k)\Big) \, 
\frac{f\Big(k_i(\vec{\Phi}_k)\Big)}
{g_{\mathrm{tot}}\Big(k_i(\vec{\Phi}_k)\Big)} \nn\\
&=&\int_0^1 \rd r_0\,  
\sum_{k=1}^M \theta(r_0-\beta_{k-1})\theta(\beta_k-r_0)
\int_0^1 \rd \vec{r} \,
\frac{f\Big(k_i(\vec{h}_k(\vec{r}\,))\Big)}
{g_{\mathrm{tot}}\Big(k_i(\vec{h}_k(\vec{r}\,))\Big)},
\end{eqnarray}
where $\beta_0=0, \beta_j=\sum_{k=1}^j \alpha_k, j=1,\ldots ,M-1$, 
$\beta_M=\sum_{k=1}^M \alpha_k=1$, and
\beq
g_{\mathrm{tot}}\Big(k_i(\vec{\Phi}_k)\Big)=
\sum_{l=1}^M \alpha_l g_l\Big(k_i(\vec{\Phi}_k)\Big),
\eeq
is the total density of the event.  

For the processes $\Pep \Pem \to 4 f$ we have between 6 and 128
different channels, for $\eeffffg$ between 14 and 928 channels.
Each channel smoothens a particular combination of propagators that
results from a characteristic Feynman diagram.  We have written
phase-space generators in a generic way for several classes of
channels determined by the chosen set of invariants $s_i, t_i$.  The
channels within one class differ in the choice of the mappings
\refeqs{eq:maps1}, \refeqf{eq:maps2}, and \refeqf{eq:nomaps} and the
order of the external particles.  We did not include special
channels for interference contributions.

The $\alpha_k$-dependence of the quantity 
\begin{equation}
W(\vec{\alpha})= \frac{1}{N} \sum_{j=1}^N [w(r_0^j,{\vec{r}}^{\,j})]^2,
\end{equation}
where $w=f/g_{\mathrm{tot}}$ is the weight assigned to the Monte Carlo point 
$(r_0^j, \vec{r}^{\,j})$ of the $j$th event,
can be exploited to minimize the expected Monte Carlo error 
\begin{equation}
\delta \bar I_n=\sqrt{\frac{W(\vec{\alpha})-\bar{I}_n^2}{N}},
\end{equation} 
with the Monte Carlo estimate of $I_n$ 
\begin{equation}
\bar I_n= \frac{1}{N} \sum_{j=1}^N w(r_0^j,\vec{r}^{\,j})
\end{equation} 
by trying to choose an optimal set of a-priori weights.
We perform the search for an optimal set of $\alpha_k$ by using an
{\it adaptive optimization} method, as described in \citere{Kl94}. 
After a certain number of generated events a new set of 
a-priori weights $\alpha_k^{\mathrm{new}}$ is calculated according to
\begin{eqnarray}
\alpha_k^{\mathrm{new}} &\propto& \alpha_k 
\sqrt{\frac{1}{N}\sum_{j=1}^N \, 
\frac{g_k\Big(k_i(\vec{h}_k({\vec{r}}^{\,j}))\Big) \, 
[w(r_0^j,{\vec{r}}^{\,j})]^2}
{g_{\mathrm{tot}}\Big(k_i(\vec{h}_k({\vec{r}}^{\,j}))\Big)}}, \qquad
\sum_{k=1}^M \alpha_k^{\mathrm{new}}=1.
\end{eqnarray}

Based on the above approach, we have written two independent Monte
Carlo programs. While the general strategy is similar, 
the programs differ in the explicit phase-space generation.

\section{Numerical results}
\label{se:numres}

If not stated otherwise we use the complex-mass scheme and the 
following parameters:
\beq
\begin{array}[b]{rlrl}
\al =& 1/128.89, & \qquad \al_s =& 0.12, \\
\MW =& 80.26\GeV,& \GW =& 2.05\GeV, \\
\MZ =& 91.1884\GeV,& \GZ =& 2.46\GeV.
\end{array}
\eeq
In the complex-mass scheme, the weak mixing angle is defined in
\refeq{complangle}, in all other schemes it 
is fixed by $\cw=\MW/\MZ$, $\sw^2=1-\cw^2$.

The energy in the centre-of-mass (CM) system of the incoming electron
and positron
is denoted by $\sqrt{s}$.
Concerning the phase-space integration, we apply
the canonical cuts of the ADLO/TH detector, 
\beq
\begin{array}[b]{rlrlrl}
\theta (l,\mathrm{beam})> & 10^\circ, & \qquad
\theta( l, l^\prime)> & 5^\circ, & \qquad 
\theta( l, q)> & 5^\circ, \\
\theta (\ga,\mathrm{beam})> & 1^\circ, &
\theta( \ga, l)> & 5^\circ, & 
\theta( \ga, q)> & 5^\circ, \\
E_\ga> & 0.1\GeV, & E_l> & 1\GeV, & E_q> & 3\GeV, \\
m(q,q')> & 5\GeV,
\end{array}
\label{eq:canonicalcuts}
\eeq
where $\theta(i,j)$ specifies the angle between the particles $i$ and
$j$ in the CM system, 
and $l$, $q$, $\ga$, and ``beam'' denote charged leptons,
quarks, photons, and the beam electrons or positrons, respectively.
The invariant mass of a quark pair $qq'$ is denoted by 
$m(q,q')$.
The cuts coincide with those defined in \citere{CERN9601mcgen},
except for the additional angular cut between charged leptons.
The canonical cuts exclude all collinear and infrared singularities 
from phase space for all processes.  

Although our helicity amplitudes and Monte Carlo programs
allow for a treatment of arbitrary polarization configurations, we
consider only unpolarized quantities in this paper.

All results are produced with $10^7$ events. 
The calculation of the cross section for
$\Pep\Pem\to\Pep\Pem\mu^+\mu^-$ requires
about 50 minutes on a DEC ALPHA workstation with 500 MHz, 
the calculation of  the cross section for
$\Pep\Pem\to\Pep\Pem\mu^+\mu^-\ga$ takes about 5 hours.
The results of our two Monte Carlo programs agree very well.
The numbers in 
parentheses in the following tables 
correspond to the statistical errors 
of the results of the Monte Carlo integrations.

\subsection{Comparison with existing results}

In order to compare our results for $\eeffff$ with Tables 6--8 
of \citere{CERN9601table},
we use the corresponding set of phase-space cuts and input parameters,
\ie the canonical cuts defined in \refeq{eq:canonicalcuts},
a CM energy of $\sqrt{s}=190\GeV$, and the parameters 
$\al=\al (2 \MW)=1/128.07$, $\al_s=0.12$, 
$\MW=80.23\GeV$, $\GW=2.0337\GeV$, $\MZ=91.1888\GeV$, and $\GZ=2.4974\GeV$.
The value of $\sw$, 
which enters the couplings, is calculated from  
$\al (2 \MW)/(2 \sw^2)=\GF \MW^2/(\pi \sqrt{2})$
with $\GF=1.16639\times 10^{-5} \GeV^{-2}$.

\begin{table}
\renewcommand{\arraystretch}{1.1}
\newdimen\digitwidth
\setbox0=\hbox{0}
\digitwidth=\wd0
\catcode`!=\active
\def!{\kern\digitwidth}
\newdimen\dotwidth
\setbox0=\hbox{$.$}
\dotwidth=\wd0
\catcode`?=\active
\def?{\kern\dotwidth}
\begin{center}
{\begin{tabular}{|c||r@{}l|r@{}l|r@{}l|}
\hline
$\si/\fb$ & 
\multicolumn{2}{c|}{$\begin{array}{c}
                     \Pep\Pem \to 4 f \\
                     \mbox{running width}
                     \end{array}$} &
\multicolumn{2}{c|}{$\begin{array}{c}
                     \Pep\Pem \to 4 f \\
                     \mbox{constant width}
                     \end{array}$} &
\multicolumn{2}{c|}{$\begin{array}{c}
                     \eeffffg \\
                     \mbox{constant width}
                     \end{array}$} 
\\\hline\hline
$\Pne \Pnebar \Pem \Pep$
&\hspace{1.15cm}$    256.7 $&$( 3)$
&\hspace{1.15cm}$    257.1 $&$( 7)$
&\hspace{0.85cm}$     89.4 $&$( 2)$
\\\hline
$\nu_\mu \mu^+ \Pem \Pnebar$
&$    227.4 $&$( 1)$
&$    227.5 $&$( 1)$
&$     79.1 $&$( 1)$
\\\hline
$\nu_\mu \bar{\nu}_\mu \mu^- \mu^+$
&$    228.7 $&$( 1)$
&$    228.8 $&$( 1)$
&$     81.0 $&$( 2)$
\\\hline
$\nu_\mu \mu^+ \tau^- \bar{\nu}_\tau$
&$   218.55 $&$( 9)$
&$   218.57 $&$( 9)$
&$     76.7 $&$( 1)$
\\\hline
$\Pem \Pep \Pem \Pep$
&$    109.1 $&$( 3)$
&$    109.4 $&$( 3)$
&$     38.8 $&$( 4)$
\\\hline
$\Pem \Pep \mu^- \mu^+$
&$    116.6 $&$( 3)$
&$    116.4 $&$( 3)$
&$     43.4 $&$( 4)$
\\\hline
$\mu^- \mu^+ \mu^- \mu^+$
&$    5.478 $&$( 5)$
&$    5.478 $&$( 5)$
&$     3.37 $&$( 1)$
\\\hline
$\mu^- \mu^+ \tau^- \tau^+$
&$    11.02 $&$( 1)$
&$    11.02 $&$( 1)$
&$     6.78 $&$( 3)$
\\\hline
$\Pem \Pep \nu_\mu \bar{\nu}_\mu $
&$   14.174 $&$( 9)$
&$   14.150 $&$( 9)$
&$     5.36 $&$( 1)$
\\\hline
$\Pne \Pnebar \mu^- \mu^+$
&$    17.78 $&$( 6)$
&$    17.73 $&$( 6)$
&$     6.63 $&$( 2)$
\\\hline
$\nu_\tau \bar{\nu}_\tau\mu^- \mu^+$
&$   10.108 $&$( 8)$
&$   10.103 $&$( 8)$
&$    4.259 $&$( 9)$
\\\hline
$\Pne \Pnebar \Pne \Pnebar$
&$    4.089 $&$( 1)$
&$    4.082 $&$( 1)$
&$   0.7278 $&$( 7)$
\\\hline
$\Pne \Pnebar \nu_\mu \bar{\nu}_\mu$
&$    8.354 $&$( 2)$
&$    8.337 $&$( 2)$
&$    1.512 $&$( 1)$
\\\hline
$\nu_\mu \bar{\nu}_\mu \nu_\mu \bar{\nu}_\mu $
&$    4.069 $&$( 1)$
&$    4.057 $&$( 1)$
&$   0.7434 $&$( 7)$
\\\hline
$\nu_\mu \bar{\nu}_\mu \nu_\tau \bar{\nu}_\tau$
&$    8.241 $&$( 2)$
&$    8.218 $&$( 2)$
&$    1.511 $&$( 1)$
\\\hline
$\Pu\, \Pdbar\, \Pem \Pnebar$
&$    693.5 $&$( 3)$
&$    693.6 $&$( 3)$
&$    220.8 $&$( 4)$
\\\hline
$\Pu\, \Pdbar\, \mu^- \bar{\nu}_\mu$
&$    666.7 $&$( 3)$
&$    666.7 $&$( 3)$
&$    214.5 $&$( 4)$
\\\hline
$\Pem \Pep \Pu\, \Pubar$
&$    86.87 $&$( 9)$
&$    86.82 $&$( 9)$
&$     32.3 $&$( 2)$
\\\hline
$\Pem \Pep \Pd\, \Pdbar$
&$    43.02 $&$( 4)$
&$    42.95 $&$( 4)$
&$    16.17 $&$( 8)$
\\\hline
$\Pu\, \Pubar\, \mu^- \mu^+$
&$    24.69 $&$( 2)$
&$    24.69 $&$( 2)$
&$    12.70 $&$( 4)$
\\\hline
$\Pd\, \Pdbar\, \mu^- \mu^+$
&$    23.73 $&$( 1)$
&$    23.73 $&$( 1)$
&$    10.43 $&$( 2)$
\\\hline
$\Pne \Pnebar \Pu\, \Pubar$
&$    24.00 $&$( 2)$
&$    23.95 $&$( 2)$
&$     6.84 $&$( 1)$
\\\hline
$\Pne \Pnebar \Pd\, \Pdbar$
&$   20.657 $&$( 8)$
&$    20.62 $&$( 1)$
&$    4.319 $&$( 6)$
\\\hline
$\Pu\, \Pubar\, \nu_\mu \bar{\nu}_\mu$
&$   21.080 $&$( 5)$
&$   21.050 $&$( 5)$
&$    6.018 $&$( 9)$
\\\hline
$\Pd\, \Pdbar\, \nu_\mu \bar{\nu}_\mu$
&$   19.863 $&$( 5)!!!$
&$   19.817 $&$( 5)!!!$
&$    4.156 $&$( 5)!!$
\\\hline
$\Pu\, \Pubar\, \Pd\, \Pdbar$
&\multicolumn{2}{c|}{$   2064.1 ( 9),   !2140.8 ( 9)$}
&\multicolumn{2}{c|}{$   2064.3 ( 9),   !!?2141 ( 1)$}
&\multicolumn{2}{c|}{$   !?615 ( 1),     !!?672 ( 1)$}
\\\hline
$\Pu\, \Pdbar\, \Ps\, \Pcbar$
&$   2015.2 $&$( 8)$
&$   2015.3 $&$( 8)$
&$      598 $&$( 1)$
\\\hline
$\Pu\, \Pubar\, \Pu\, \Pubar$
&\multicolumn{2}{c|}{$   25.738 ( 7),    !!71.28 ( 4)$}
&\multicolumn{2}{c|}{$   25.721 ( 7),    !!71.30 ( 4)$}
&\multicolumn{2}{c|}{$    !9.78 ( 2),    !! 42.1 ( 1)$}
\\\hline
$\Pd\, \Pdbar\, \Pd\, \Pdbar$
&\multicolumn{2}{c|}{$   23.494 ( 6),    !!51.35 ( 3)$}
&\multicolumn{2}{c|}{$   23.448 ( 6),    !!51.32 ( 3)$}
&\multicolumn{2}{c|}{$    5.527 ( 7),    !28.68 ( 4)$}
\\\hline
$\Pu\, \Pubar\, \Pc\, \Pcbar$
&\multicolumn{2}{c|}{$    !51.61 ( 1),   !144.72 ( 9)$}
&\multicolumn{2}{c|}{$    !51.57 ( 1),   !144.75 ( 9)$}
&\multicolumn{2}{c|}{$    19.61 ( 4),    !! 86.1 ( 2)$}
\\\hline
$\Pu\, \Pubar\, \Ps\, \Psbar$
&\multicolumn{2}{c|}{$    !49.68 ( 1),   !126.52 ( 8)$}
&\multicolumn{2}{c|}{$    !49.62 ( 1),   !126.52 ( 8)$}
&\multicolumn{2}{c|}{$    15.17 ( 2),    !! 75.1 ( 2)$}
\\\hline
$\Pd\, \Pdbar\, \Ps\, \Psbar$
&\multicolumn{2}{c|}{$    !47.13 ( 1),   !104.79 ( 6)$}
&\multicolumn{2}{c|}{$    !47.02 ( 1),   !104.74 ( 6)$}
&\multicolumn{2}{c|}{$    11.10 ( 2),    !! 59.2 ( 1)$}
\\\hline
\end{tabular}}
\end{center}
\caption[]{Integrated cross sections for all representative  processes
  $\eeffff$ with running widths and constant widths and for the
  corresponding processes $\eeffffg$ with constant widths.
  If two numbers are given, the first  results from
pure electroweak diagrams and the second involves in addition 
gluon-exchange contributions.}
\label{ta:yellowreport}
\end{table}
In \refta{ta:yellowreport}, we list the integrated cross sections for
various processes $\eeffff$ with running widths and
constant widths,
and for the corresponding processes $\eeffffg$ 
with constant widths. For processes involving gluon-exchange
diagrams we give the cross sections resulting from the purely
electroweak diagrams and those including the gluon-exchange
contributions. The latter results include also the interference terms
between purely electroweak and gluon-exchange diagrams.
In \refta{ta:yellowreport} we provide a complete list of processes
for vanishing fermion masses. All processes $\eeffff(\ga)$ not explicitly
listed are equivalent to one of the given processes.

For NC processes $\eeffff$ 
with four neutrinos or four quarks in the final state
we find small deviations of roughly $0.2\%$ between the results with
constant and running widths. Assuming that a running width has been
used in \citere{CERN9601table}, we find very good agreement.

Unfortunately we cannot compare with most of the publications
\cite{vO94,vO96,Fu94,Ca97} for the bremsstrahlung processes
$\eeffffg$. In those papers, either the cuts are not (completely) specified,
or collinear photon emission is not excluded, and the corresponding
fermion-mass effects are taken into account.  Note that the
contributions of collinear photons dominate the results given there.

We have compared our results with the ones given in \citeres{Ae91,Ae91a},
where the total cross sections for $\eeffffg$ have been calculated for
the purely leptonic and the semi-leptonic final states. As done in
\citeres{Ae91,Ae91a} only diagrams involving two resonant \PW~bosons have
been taken into account
for this comparison. Table~\ref{ta:aepplitable} contains our
results corresponding to \refta{ta:aepplitable} of \citere{Ae91}.
Based on \citeres{Ae91,Ae91a},
we have chosen $\sqrt{s}=200\GeV$ and the input parameters
$\al=1/137.03599$, $\MW=80.9\GeV$, $\GW=2.14\GeV$, $\MZ=91.16\GeV$, $\GZ=2.46\GeV$,
$\sw$ 
obtained from $\al/(2 \sw^2)=\GF \MW^2/(\pi \sqrt{2})$ with
$\GF=1.16637\times 10^{-5} \GeV^{-2}$, and constant gauge-boson
widths.  
The energy of the photon is
required to be larger than $E_{\ga,\min}$,
and the angle between the
photon and any charged fermion must be larger than
$\theta_{\ga,\min}$. A maximal photon energy is required,
$E_\ga<60\GeV$, in order to exclude contributions from the $\PZ$
resonance. Our results 
are consistent with those of \citeres{Ae91,Ae91a} within
the statistical error of 1\% given there. In some cases we find
deviations of 2\%.%
\footnote{Note that the input specified in \citeres{Ae91,Ae91a} is not
  completely clear even if the information of both publications is
  combined.}
\begin{table}
\newdimen\digitwidth
\setbox0=\hbox{0}
\digitwidth=\wd0
\catcode`!=\active
\def!{\kern\digitwidth}
\begin{center}
{\begin{tabular}{|l|c|r@{}l|r@{}l|r@{}l|r@{}l|}
\hline
\multicolumn{2}{|r|}{$E_{\ga,\min}=$} &
\multicolumn{2}{c|}{$1\GeV$} &
\multicolumn{2}{c|}{$5\GeV$} & 
\multicolumn{2}{c|}{$10\GeV$} & 
\multicolumn{2}{c|}{$15\GeV$} \\
\hline
&$\theta_{\ga,\min}$& 
\multicolumn{8}{c|}{$\si/\mathrm{fb}$}  \\\hline\hline
&$ !1^\circ  $
&$    53.54 $&$( 8)$
&$    27.57 $&$( 3)$
&$    16.96 $&$( 2)$
&$    11.22 $&$( 2)$
\\\cline{2-10}
leptonic
&$!5^\circ  $
&$    32.65 $&$( 4)$
&$    16.98 $&$( 3)$
&$    10.48 $&$( 2)$
&$     6.94 $&$( 1)$
\\\cline{2-10}
process
&$10^\circ $
&$    23.48 $&$( 3)$
&$    12.30 $&$( 2)$
&$     7.61 $&$( 2)$
&$     5.04 $&$( 1)$
\\\cline{2-10}
&$15^\circ $
&$    18.03 $&$( 2)$
&$     9.51 $&$( 2)$
&$     5.90 $&$( 1)$
&$     3.90 $&$( 1)$
\\\hline\hline
&$!1^\circ  $
&$    141.9 $&$( 2)$
&$    71.90 $&$( 8)$
&$    43.56 $&$( 5)$
&$    28.26 $&$( 4)$
\\\cline{2-10}
semi-leptonic
&$!5^\circ  $
&$     86.8 $&$( 1)$
&$    44.25 $&$( 6)$
&$    26.78 $&$( 4)$
&$    17.40 $&$( 3)$
\\\cline{2-10}
process
&$10^\circ $
&$    62.29 $&$( 7)$
&$    31.92 $&$( 5)$
&$    19.40 $&$( 4)$
&$    12.61 $&$( 3)$
\\\cline{2-10}
&$15^\circ $
&$    47.42 $&$( 5)$
&$    24.50 $&$( 4)$
&$    14.97 $&$( 3)$
&$     9.77 $&$( 2)$
\\\hline
\end{tabular}}
\end{center}
\caption[]{Comparison with Table 2 of \citere{Ae91}:
Cross sections resulting from diagrams involving two resonant W~bosons
for purely 
leptonic and semi-leptonic final states  and several photon separation
cuts
}
\label{ta:aepplitable}
\end{table}

\subsection{Comparison of finite-width schemes}

{}As discussed in \citeres{bhf2,FLscheme},
particular care has to be
taken when implementing the finite gauge-boson widths.  Differences
between results obtained with running or constant widths can already
be seen in \refta{ta:yellowreport}, where a typical LEP2 energy is
considered.
\begin{table}
\begin{center}
{\begin{tabular}{|c|c|r@{}l|r@{}l|r@{}l|r@{}l|}
\hline
\multicolumn{1}{|c|}{$\si/\fb$} &
\multicolumn{1}{r|}{$\sqrt{s}=$} & 
\multicolumn{2}{c|}{$189\GeV$} & 
\multicolumn{2}{c|}{$500\GeV$} &
\multicolumn{2}{c|}{$2\TeV$} & 
\multicolumn{2}{c|}{$10\TeV$}
\\\hline\hline
& constant width
&$    703.5 $&$( 3)$
&$    237.4 $&$( 1)$
&$    13.99 $&$( 2)$
&$    0.624 $&$( 3)$
\\\cline{2-10}
$\Pep \Pem \to \Pu\, \Pdbar\, \mu^- \bar{\nu}_\mu $
& running width
&$    703.4 $&$( 3)$
&$    238.9 $&$( 1)$
&$    34.39 $&$( 3)$
&$    498.8 $&$( 1)$
\\\cline{2-10}
& complex-mass scheme
&$    703.1 $&$( 3)$
&$    237.3 $&$( 1)$
&$    13.98 $&$( 2)$
&$    0.624 $&$( 3)$
\\\hline\hline
& constant width
&$    224.0 $&$( 4)$
&$     83.4 $&$( 3)$
&$     6.98 $&$( 5)$
&$    0.457 $&$( 6)$
\\\cline{2-10}
$\Pep \Pem \to \Pu\, \Pdbar\, \mu^- \bar{\nu}_\mu \,\ga$
& running width
&$    224.6 $&$( 4)$
&$     84.2 $&$( 3)$
&$     19.2 $&$( 1)$
&$      368 $&$( 6)$
\\\cline{2-10}
& complex-mass scheme
&$    223.9 $&$( 4)$
&$     83.3 $&$( 3)$
&$     6.98 $&$( 5)$
&$    0.460 $&$( 6)$
\\\hline\hline
& constant width
&$    730.2 $&$( 3)$
&$    395.3 $&$( 2)$
&$    211.0 $&$( 2)$
&$    32.38 $&$( 6)$
\\\cline{2-10}
$\Pep \Pem \to \Pu\, \Pdbar\, \Pem \Pnebar$
& running width
&$    729.8 $&$( 3)$
&$    396.9 $&$( 2)$
&$    231.5 $&$( 2)$
&$    530.2 $&$( 6)$
\\\cline{2-10}
& complex-mass scheme
&$    729.8 $&$( 3)$
&$    395.1 $&$( 2)$
&$    210.9 $&$( 2)$
&$    32.37 $&$( 6)$
\\\hline\hline
& constant width
&$    230.0 $&$( 4)$
&$    136.5 $&$( 5)$
&$     84.0 $&$( 7)$
&$     16.8 $&$( 5)$
\\\cline{2-10}
$\Pep \Pem \to \Pu\, \Pdbar\, \Pe^- \Pnebar \,\ga $
& running width
&$    230.6 $&$( 4)$
&$    137.3 $&$( 5)$
&$     95.7 $&$( 7)$
&$      379 $&$( 6)$
\\\cline{2-10}
& complex-mass scheme
&$    229.9 $&$( 4)$
&$    136.4 $&$( 5)$
&$     84.1 $&$( 6)$
&$     16.8 $&$( 5)$
\\\hline
\end{tabular}}
\end{center}
\caption[]{Comparison of different width schemes for several processes 
and energies}
\label{ta:width}
\end{table}
In \refta{ta:width} we compare predictions for integrated cross
sections obtained 
by using a constant width, a running width, or the
complex-mass scheme for several energies. We consider
two semi-leptonic final states for $\eeffff(\ga)$.
The numbers show that the constant width and the complex-mass scheme
yield the same results within the statistical accuracy for
$\eeffff$ and
$\eeffffg$. In contrast, the results with the running width produce
totally wrong results for high energies.  The difference of the
running width with respect to the other implementations of the finite
width is up to $1\%$ already for $500\GeV$. Thus, the running width
should not be used for linear-collider energies.  As already stated
above, our default treatment of the finite width is the complex-mass
scheme in the following.

\subsection{Survey of photon-energy spectra}
\label{se:photonspectra}

\begin{figure}
\centerline{
\setlength{\unitlength}{1cm}
\begin{picture}(7.2,7)
\put(0,0){\includegraphics{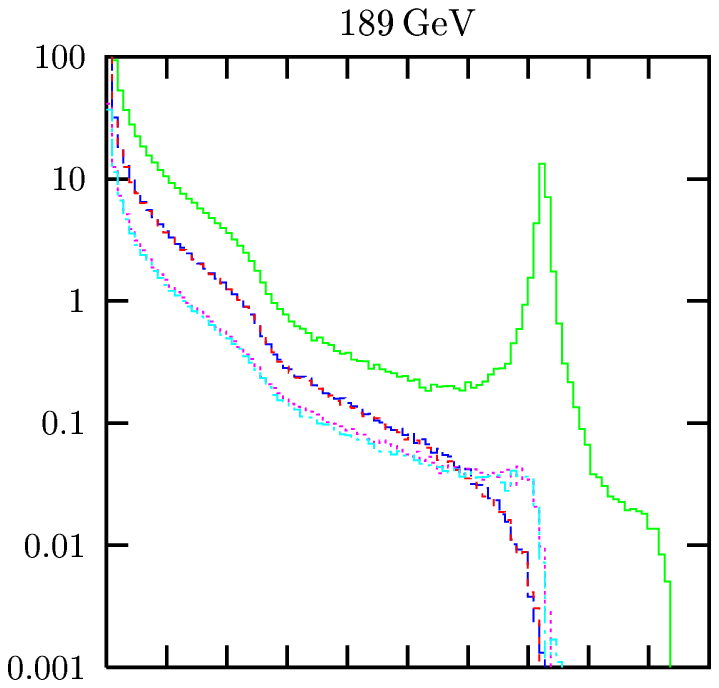}}
\put(-0.3,4.9){\makebox(1,1)[c]{$\frac{\rd \si}{\rd E_\ga}/
                              \frac{\fb}\GeV$}}
\end{picture}
\begin{picture}(7.2,7)
\put(0,0){\includegraphics{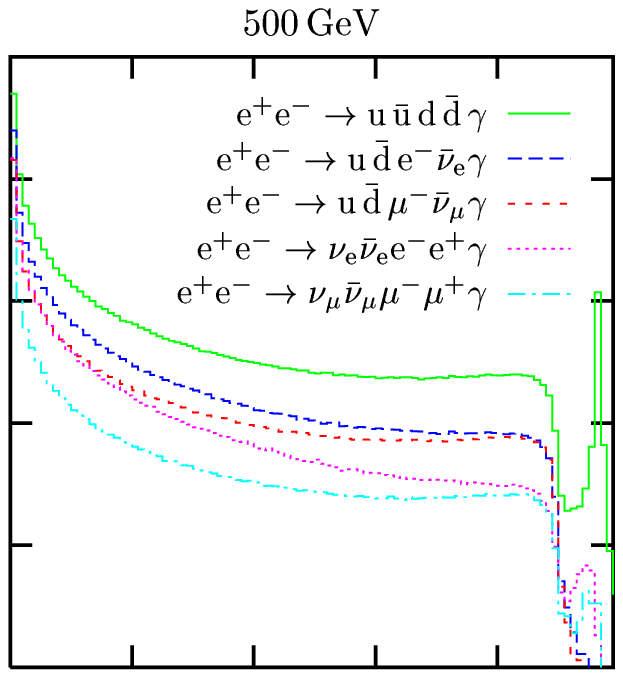}}
\end{picture}
}
\centerline{
\setlength{\unitlength}{1cm}
\begin{picture}(7.2,7)
\put(0,0){\includegraphics{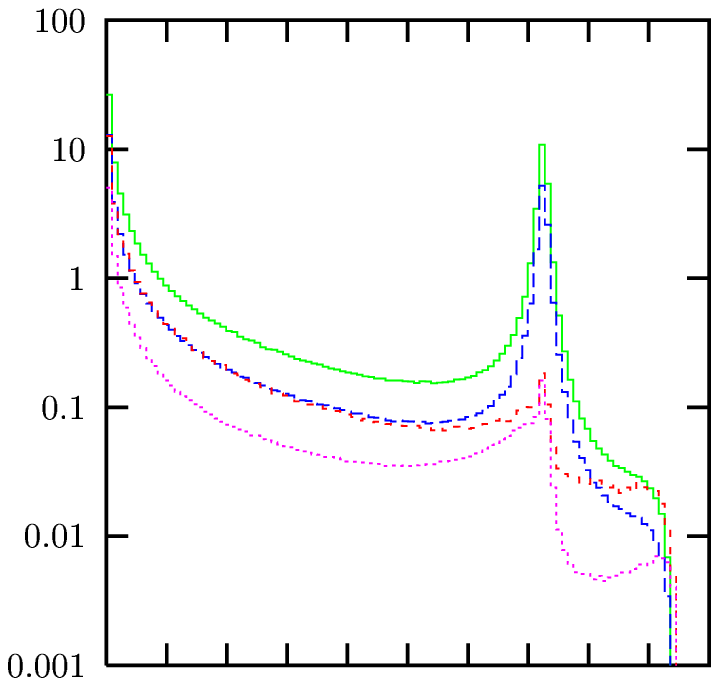}}
\put(-0.3,5.15){\makebox(1,1)[c]{$\frac{\rd \si}{\rd E_\ga}/
                              \frac{\fb}\GeV$}}
\end{picture}
\begin{picture}(7.2,7)
\put(0,0){\includegraphics{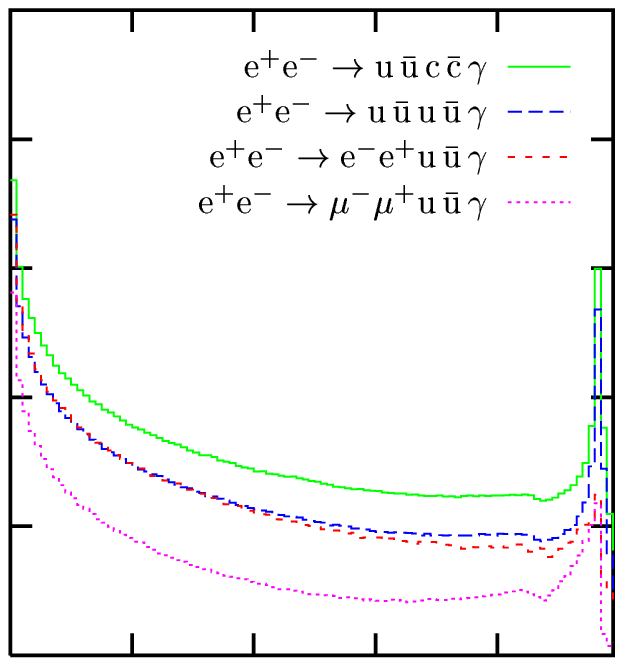}}
\end{picture}
}
\centerline{
\setlength{\unitlength}{1cm}
\begin{picture}(7.2,7)
\put(0,0){\includegraphics{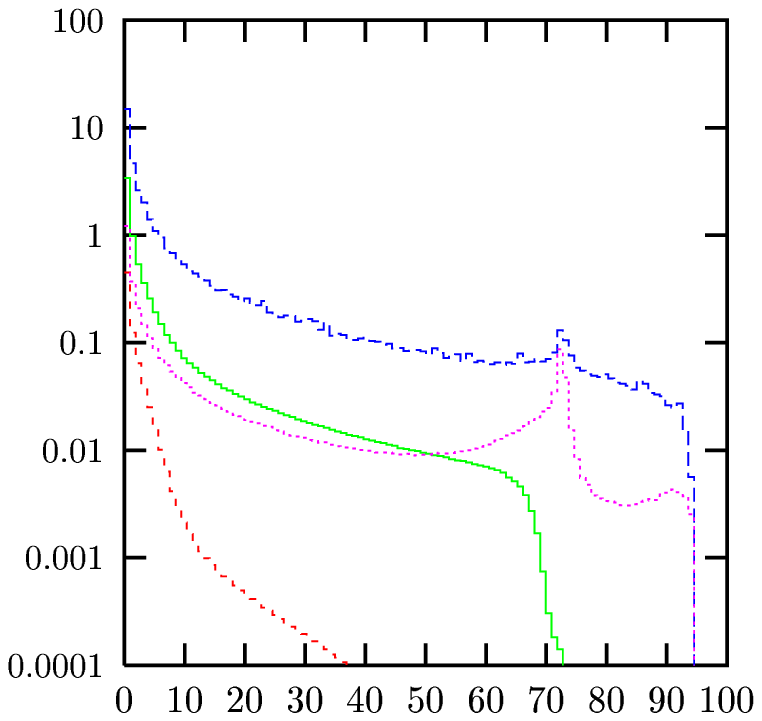}}
\put(-0.3,5.5){\makebox(1,1)[c]{$\frac{\rd \si}{\rd E_\ga}/
                              \frac{\fb}\GeV$}}
\put(4.5,-0.3){\makebox(1,1)[cc]{{$E_\ga/\GeV$}}}
\end{picture}
\begin{picture}(7.2,7)
\put(0,0){\includegraphics{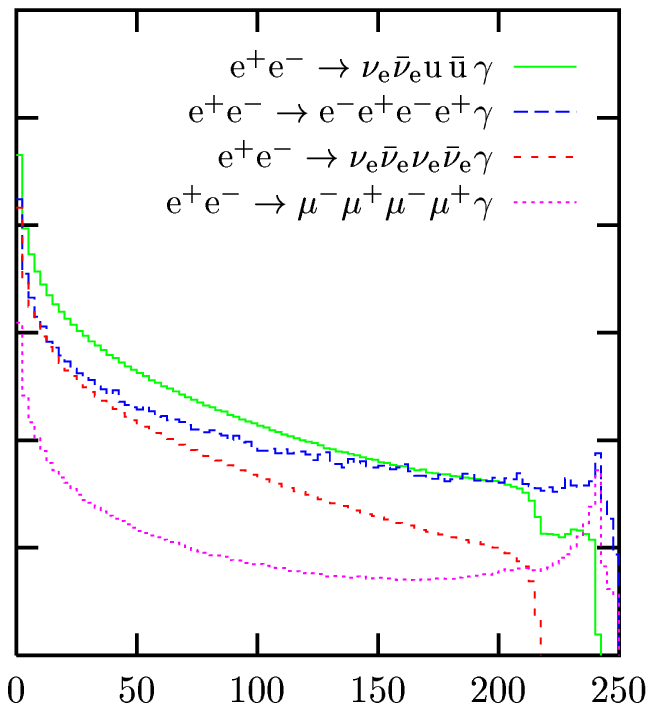}}
\put(3.5,-0.3){\makebox(1,1)[cc]{{$E_\ga/\GeV$}}}
\end{picture}
}
\caption[]{Photon-energy spectra for several processes and 
for  $\sqrt{s}=189\GeV$ and $500\GeV$}
\label{fi:photonspectra}
\end{figure}

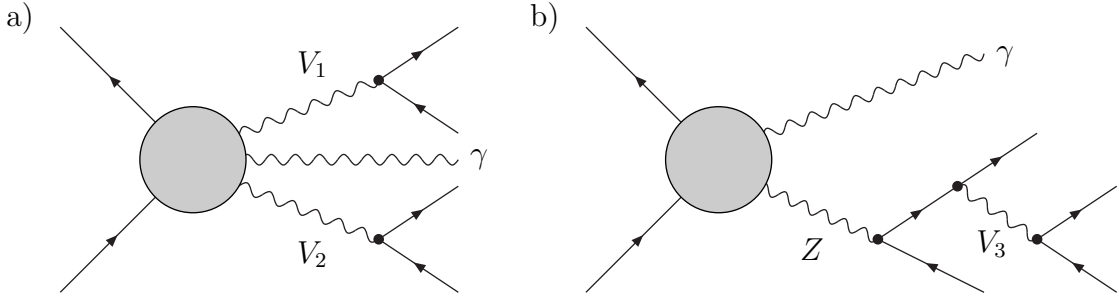
\begin{figure}
{
\begin{center}
\begin{picture}(200,120)(-5,0)
\ArrowLine(0,10)(40,50)
\ArrowLine(40,70)(0,110)
\Photon(50,60)(150,60){2}{11}
\Photon(50,63)(120,90){2}{8}
\Photon(50,57)(120,30){2}{8}
\Vertex(120,90){2}
\Vertex(120,30){2}
\ArrowLine(120,90)(150,110)
\ArrowLine(150,70)(120,90)
\ArrowLine(120,30)(150,50)
\ArrowLine(150,10)(120,30)
\GCirc(50,60){20}{0.8}
\put(95,24){\makebox(1,1)[c]{$V_2$}}
\put(95,96){\makebox(1,1)[c]{$V_1$}}
\put(158,60){\makebox(1,1)[c]{$\gamma$}}
\Text(-10,120)[rt]{a)}
\end{picture}
\begin{picture}(200,120)
\ArrowLine(0,10)(40,50)
\ArrowLine(40,70)(0,110)
\Photon(50,63)(150,100){2}{12}
\Photon(50,57)(110,30){2}{8}
\Vertex(110,30){2}
\ArrowLine(110,30)(140,50)\ArrowLine(140,50)(170,70)
\ArrowLine(150,10)(110,30)
\Vertex(140,50){2}
\Photon(140,50)(170,30){2}{4}
\Vertex(170,30){2}
\ArrowLine(170,30)(200,50)
\ArrowLine(200,10)(170,30)
\GCirc(50,60){20}{0.8}
\put(85,26){\makebox(1,1)[c]{$Z$}}
\put(153,28){\makebox(1,1)[c]{$V_3$}}
\put(158,99){\makebox(1,1)[c]{$\gamma$}}
\Text(-10,120)[rt]{b)}
\end{picture}
\end{center}
}
\caption[]{Diagrams for important subprocesses, where
  $V_1,V_2=\PW,\PZ,\ga$, and $V_3=\ga,\Pg$ }
\label{fi:resonance}
\end{figure}

In \reffi{fi:photonspectra} we show the photon-energy spectra of several
processes for the typical LEP2 energy of $189\GeV$ and a possible
linear-collider energy of $500\GeV$.
The upper plots contain CC and CC/NC processes, the plots in the
middle and the lower plots contain NC processes.
Several spectra show threshold or peaking structures. These structures
are caused by diagrams in which the
photon is emitted from the initial state. The two important classes of
diagrams are  shown in \reffi{fi:resonance}.

The first class, shown in \reffi{fi:resonance}a, corresponds
to triple-gauge-boson-production subprocesses which yield dominant
contributions as long as the two virtual gauge bosons $V_1$ and $V_2$
can become simultaneously resonant. If the real photon takes the
energy $E_\ga$, defined
in the CM system, only the energy $\sqrt{s'}$, with
\beq
s' = s-2 \sqrt{s}\, E_\ga,
\eeq
is available for the production of the gauge-boson pair $V_1V_2$.
If at least one of the gauge bosons is massive, 
and if the photon becomes too hard, 
the two gauge bosons cannot be produced
on shell anymore, so that the spectrum falls off for $E_\gamma$ above
the corresponding threshold $E_\gamma^{V_1 V_2}$.
Using the threshold condition for the on-shell production of the
$V_1V_2$ pair,
\beq
\sqrt{s'}>M_{V_1}+M_{V_2},
\eeq
the value of $E_\gamma^{V_1 V_2}$ is determined by 
\beq
E_\gamma^{V_1 V_2} = \frac{s-(M_{V_1}+M_{V_2})^2}{2\sqrt{s}}.
\eeq
The values of the photon energies 
that cause such thresholds
can be found in \refta{ta:resonances}.
\begin{table}
\renewcommand{\arraystretch}{1.1}
\begin{center}
{\begin{tabular}{|c|c|c|c|c|c|c|c|c|}
\hline
$\sqrt{s}/\GeV$ &
\multicolumn{4}{c|}{$189$} &
\multicolumn{4}{c|}{$500$} 
\\ \hline \hline
$V_1 V_2$ & 
$\PW \PW$ & $\PZ \PZ$ & $\ga \PZ$ & $\ga \ga$ &
$\PW \PW$ & $\PZ \PZ$ & $\ga \PZ$ & $\ga \ga$ 
\\ \hline
$E_\ga^{V_1 V_2}/\GeV$ & 
$26.3$ & $6.5$ & $72.5$& $94.5$ &
$224$ & $217$ & $242$ & $250$
\\ \hline
\end{tabular}}
\end{center}
\caption[]{Photon energies $E^{V_1 V_2}_\ga$ corresponding to thresholds}
\label{ta:resonances}
\end{table}
The value $E^{\gamma\gamma}_\ga$ corresponds to the 
upper endpoint of the photon-energy  spectrum, which is given by the beam
energy $\sqrt{s}/2$.
Since $\sqrt{s'}$ is fully determined by $s$ and
$E_\ga$, the contribution of the $V_1 V_2$-production subprocess to the 
$E_\ga$ spectrum qualitatively follows the energy
dependence of the total cross section for $V_1 V_2$ production (\cf
\citere{CERN9601table}, Fig.~1) above the corresponding thresholds.  
The cross sections for $\ga\ga$ and $\ga\PZ$ production strongly
increase with decreasing energy, while the ones for $\PZ\PZ$ and
$\PW\PW$ production are comparably flat.
Thus, the $\ga\ga$ and $\ga\PZ$-production subprocesses introduce contributions
in the photon-energy  spectra with resonance-like structures,
whereas the ones with $\PZ\PZ$ or $\PW\PW$ pairs yield 
edges.

The second class of important diagrams, shown in 
\reffi{fi:resonance}b corresponds to the production of a photon and a
resonant \PZ~boson that decays into four fermions. These diagrams are
important 
if the gauge boson $V_3$ 
is also resonant, \ie a photon or a gluon with small
invariant mass. In this case, the kinematics fixes the energy of the
real photon to
\beq
E_\ga = E_{\ga}^{\ga\PZ}=\frac{s-\MZ^2}{2\sqrt{s}},
\eeq
which corresponds to the $\ga\PZ$ threshold  in
\refta{ta:resonances}. This subprocess gives rise to resonance
structures at $E_{\ga}^{\ga\PZ}$, which are even enhanced by
$\al_{\mathrm{s}}/\al$ in the presence of gluon exchange.

In the photon-energy spectra of \reffi{fi:photonspectra} all these
threshold and resonance effects are visible.  The effect of the $\ga
\PZ$ peak can be nicely seen in different photon-energy spectra, in
particular in those where gluon-exchange diagrams contribute (\cf also
\reffi{fi:QCD}).  The effect of the WW threshold is present in the
upper two plots of \reffi{fi:photonspectra}. In the plot for
$\sqrt{s}=189\GeV$ the threshold for single W production causes the
steep drop of the spectrum for the pure CC processes above $70\GeV$.
Note that the CC cross sections are an order of magnitude larger than
the NC cross sections if the WW channel is open.  The ZZ threshold is
visible in the middle and lower plots for $\sqrt{s}=500\GeV$. The
$\ga$Z threshold (resulting from the graphs of \reffi{fi:resonance}a)
is superimposed on the $\ga \PZ$ peak (resulting from the graphs of
\reffi{fi:resonance}b) and therefore best recognizable in those
channels where the $\ga \PZ$ peak is absent or suppressed, \ie where a
neutrino pair is present in the final state or where at least no
gluon-exchange diagrams contribute.  Processes with four neutrinos in
the final state do not involve photonic diagrams and are therefore
small above the ZZ threshold.  The effects of the
triple-photon-production subprocess appear as a tendency of some
photon-energy spectra to increase near the maximal value of $E_\ga$
for two charged fermion--antifermion pairs in the final state.


\subsection{Triple-gauge-boson-production subprocesses}

\begin{figure}
\centerline{
\setlength{\unitlength}{1.1cm}
\begin{picture}(14.5,6.3)
\put(0.8,0){\includegraphics{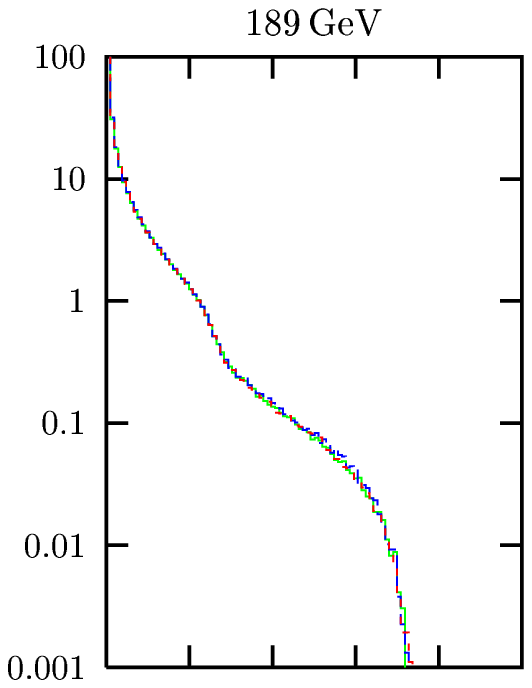}}
\put(5.1,0){\includegraphics{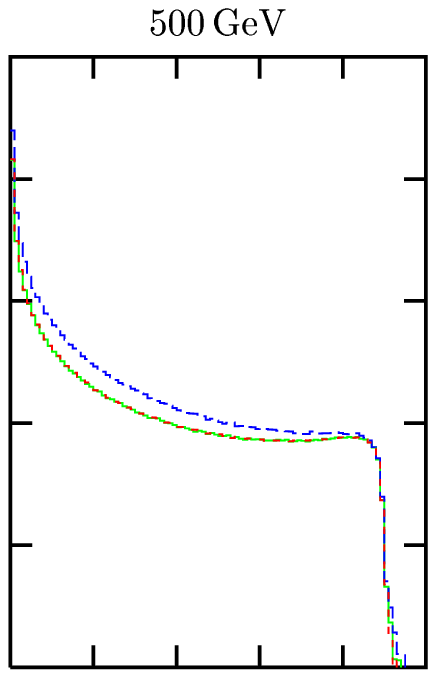}}
\put(9.4,0){\includegraphics{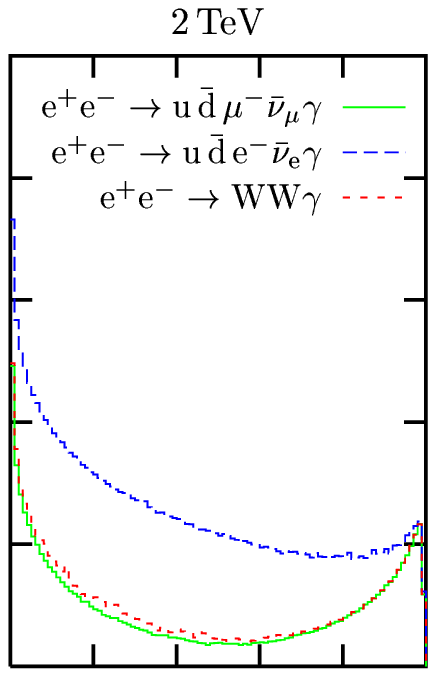}}
\put(0,5.5){\makebox(1,1)[c]{$\frac{\rd \si}{\rd E_\ga}/
                                 \frac{\fb}\GeV$}}
\end{picture}
}
\centerline{
\setlength{\unitlength}{1.1cm}
\begin{picture}(14.5,6.3)
\put(0.8,0){\includegraphics{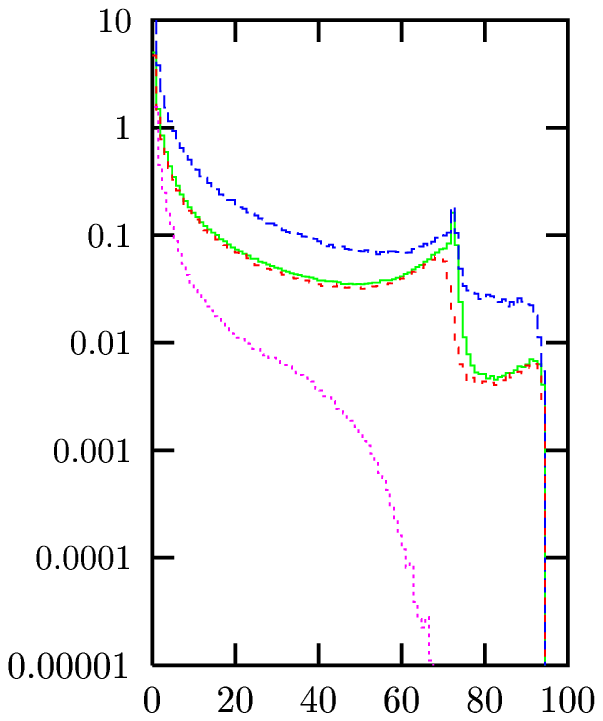}}
\put(5.1,0){\includegraphics{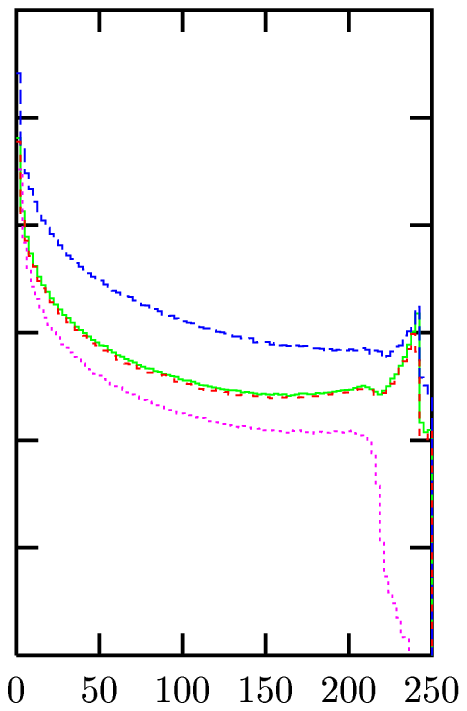}}
\put(9.4,0){\includegraphics{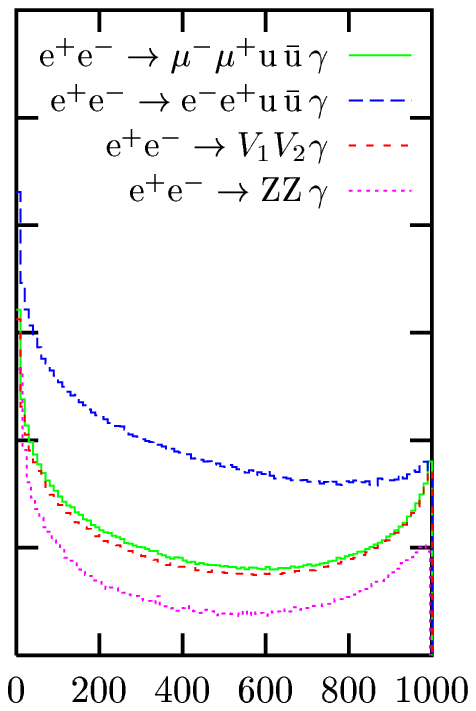}}
\put(0,5.95){\makebox(1,1)[c]{$\frac{\rd \si}{\rd E_\ga}/
                                 \frac{\fb}\GeV$}}
\put(7.5,-0.3){\makebox(1,1)[cc]{{$E_\ga/\GeV$}}}
\end{picture}
}
\caption[]{Photon-energy spectra resulting from the
  triple-gauge-boson-production subprocesses compared to those
  resulting from all diagrams ($V_1V_2$ includes ZZ, $\ga$Z, and
  $\ga\ga$)}
\label{fi:signal}
\end{figure}

In \reffi{fi:signal} we compare predictions that are based on the full
set of diagrams with those that include only the graphs associated
with the triple-gauge-boson-production subprocesses, i.e.\ the graphs
in \reffi{fi:resonance}a.  In addition we consider the contributions
of the $\PZ\PZ\ga$-production subprocess alone.
For CC processes, the photon-energy spectra resulting from the
$\PWp\PWm\ga$-production subprocess are close to those resulting from
all diagrams at LEP2 energies, but large differences are found for
higher energies and $\Pe^\pm$ in the final state. Note that the
spectra are shown on a logarithmic scale.  Even at LEP2 energies the
differences between the predictions for different final states may be
important, as can be seen, for instance, in \refta{ta:yellowreport} by
comparing the cross sections of $\Pep\Pem\to
\Pu\,\Pdbar\,\mu^-\bar\nu_\mu\ga$ and $\Pep\Pem\to
\Pu\,\Pdbar\,\Pem\Pnebar\ga$.  In the case of NC processes, already
for $189\GeV$ the contributions from $\PZ\PZ\ga$, $\PZ\ga\ga$, and
$\ga\ga\ga$ production are not sufficient: in the vicinity of the
$\ga\PZ$ peak sizeable contributions result from the
$\ga\PZ$-production subprocess (\reffi{fi:resonance}b) even for the
$\mu^+\mu^-\Pu\,\Pubar\,\ga$ final state.  For
$\Pep\Pem\to\Pem\Pep\Pu\,\Pubar\,\ga$ other diagrams become dominating
everywhere.  The contribution of $\PZ\PZ\ga$ production is always
small and could only be enhanced by invariant-mass cuts.  Note that
the triple-gauge-boson-production diagrams form a gauge-invariant
subset for NC processes, while this is not the case for CC processes.

\subsection{Relevance of gluon-exchange contributions}
\label{se:QCD}

In the analytical calculation of the matrix elements for
$\eeffff(\ga)$ in \refse{se:anres} we have seen that NC processes
with four quarks in the final state involve, besides purely electroweak,
 also gluon-exchange diagrams. Table \ref{ta:QCD}
illustrates the impact of these diagrams on the integrated cross
sections for a CM energy of $500\GeV$.
\begin{table}
\begin{center}
{\begin{tabular}{|l|r@{}l|r@{}l|r@{}l|r@{}l|}
\hline
\multicolumn{1}{|c|}{$\si/\fb$} & 
\multicolumn{2}{c|}{ew and gluon} & 
\multicolumn{2}{c|}{purely ew} & 
\multicolumn{2}{c|}{gluon} &
\multicolumn{2}{c|}{interference} 
\\\hline\hline
$\Pep \Pem \to \Pu\, \Pubar\, \Pc\, \Pcbar$
&\hspace{0.5cm}$    52.98 $&$( 4)$
&$   21.560 $&$( 6)$
&\hspace{0.2cm}$    31.38 $&$( 3)$
&\hspace{0.35cm}$     0.04 $&$( 5)$
\\\hline
$\Pep \Pem \to \Pu\, \Pubar\, \Pc\, \Pcbar\, \ga$
&$     29.8 $&$( 1)$
&$    10.38 $&$( 4)$
&$     19.6 $&$( 1)$
&$     -0.1 $&$( 1)$
\\\hline
$\Pep \Pem \to \Pu\, \Pubar \,\Pu\, \Pubar$
&$    26.25 $&$( 2)$
&$   10.765 $&$( 3)$
&$    15.34 $&$( 1)$
&$     0.14 $&$( 2)$
\\\hline
$\Pep \Pem \to \Pu\, \Pubar\, \Pu\, \Pubar\, \ga$
&$    14.83 $&$( 7)$
&$     5.16 $&$( 2)$
&$     9.52 $&$( 5)$
&$     0.15 $&$( 9)$
\\\hline
$\Pep \Pem \to \Pd\, \Pdbar\, \Pu\, \Pubar$
&$    901.2 $&$( 6)$
&$    876.4 $&$( 5)$
&$    24.24 $&$( 2)$
&$      0.6 $&$( 8)$
\\\hline
$\Pep \Pem \to \Pd\, \Pdbar\, \Pu\, \Pubar\, \ga$
&$      290 $&$( 1)$
&$      275 $&$( 1)$
&$    14.82 $&$( 8)$
&$        0 $&$( 1)$
\\\hline
\end{tabular}}
\end{center}
\caption[]{Full lowest order cross section (ew and gluon) and
  contributions of purely electroweak diagrams (ew), of
  gluon-exchange diagrams (gluon), and their interference for $500\GeV$}
\label{ta:QCD}
\end{table}
The results for the interference are obtained by subtracting the
purely electroweak and the gluon contribution from the total cross
section.  For pure NC processes the contributions of gluon-exchange
diagrams dominate over the purely electroweak graphs. This can be
understood from the fact that the gluon-exchange diagrams are enhanced
by the strong coupling constant, and, as discussed in
\refse{se:photonspectra}, that the diagrams with gluons replaced by
photons yield a sizeable contribution to the cross section.  For the
mixed CC/NC processes the purely electroweak diagrams dominate the
cross section. Here, the contributions from the
$\PWp\PWm\ga$-production subprocess are large compared to all other
diagrams, even if the latter are enhanced by the strong coupling.  At
$500\GeV$ the gluon-exchange diagrams contribute to the cross section
at the level of several per cent. The interference contributions are
relatively small.  As discussed at the end of \refse{se:hadfinstat},
this is due to the fact that interfering electroweak and
gluon-exchange diagrams involve different resonances. Note that the
interference vanishes for $\Pep \Pem \to
\Pu\,\Pubar\,\Pc\,\Pcbar\,\ga$, and the corresponding numbers in
\refta{ta:QCD} are only due to the Monte Carlo integration error.

\begin{figure}
\centerline{
\setlength{\unitlength}{1cm}
\begin{picture}(7.2,8)
\put(0,0){\includegraphics{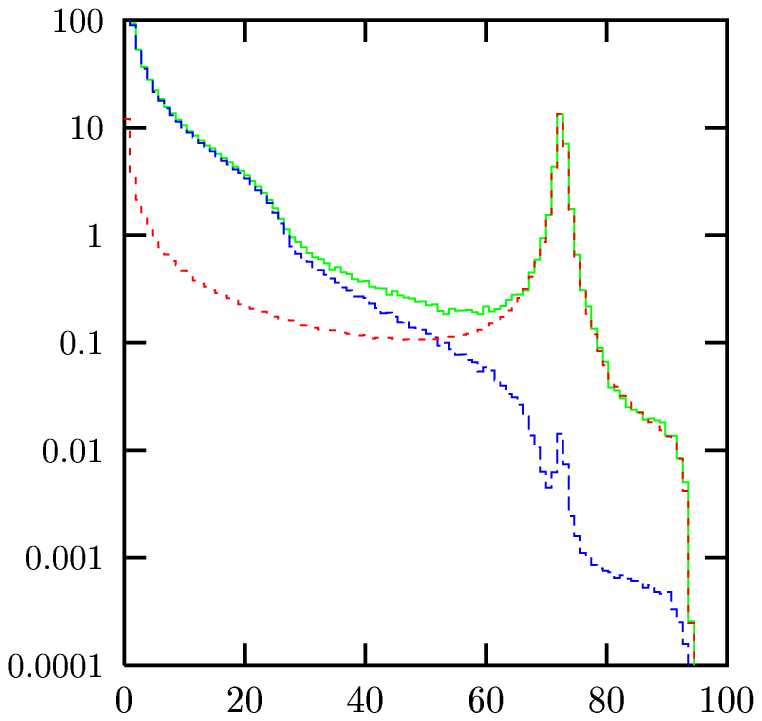}}
\put(-0.3,5.2){\makebox(1,1)[c]{$\frac{\rd \si}{\rd E_\ga}/
                              \frac{\fb}\GeV$}}
\put(4.5,-0.3){\makebox(1,1)[cc]{{$E_\ga/\GeV$}}}
\put(5.2,6.7){\makebox(1,1)[cc]{\small
        {$\Pep \Pem \to \Pu\,\Pubar\,\Pd\,\Pdbar\,\ga$}}}
\end{picture}
\begin{picture}(7.2,8)
\put(0,0){\includegraphics{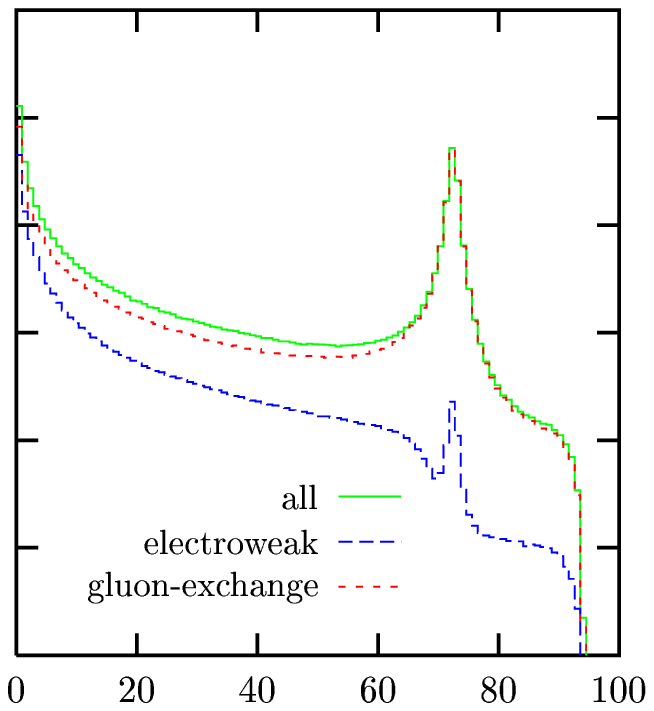}}
\put(3.5,-0.3){\makebox(1,1)[cc]{{$E_\ga/\GeV$}}}
\put(4.5,6.7){\makebox(1,1)[cc]{
        {\small$\Pep \Pem \to \Pu\,\Pubar\,\Pu\,\Pubar\,\ga$}}}
\end{picture}
}
\caption[]{Electroweak and gluon-exchange contributions to the
  photon-energy spectra for $\Pep \Pem \to
  \Pu\,\Pubar\,\Pd\,\Pdbar\,\ga$ and $\Pep \Pem \to
  \Pu\,\Pubar\,\Pu\,\Pubar\,\ga$ at $\sqrt{s}=189\GeV$}
\label{fi:QCD}
\end{figure}

In \reffi{fi:QCD} we show the photon-energy spectra for the processes
$\Pep \Pem \to \Pu\,\Pubar\,\Pd\,\Pdbar\,\ga$ and $\Pep \Pem \to
\Pu\,\Pubar\,\Pu\,\Pubar\,\ga$ together with the separate
contributions from purely electroweak and gluon-exchange diagrams.
The pure electroweak contributions are similar to the ones for 
$\Pep \Pem \to \Pu\,\Pdbar\,\mu^-\bar\nu_\mu \,\ga$ and $\Pep \Pem \to
\Pem \Pep \Pu\, \Pubar \,\ga$ in \reffi{fi:photonspectra}.  For the NC
process $\Pep\Pem \to \Pu\,\Pubar\,\Pu\,\Pubar\,\ga$, the
photon-energy spectrum is dominated by the gluon-exchange
contribution, which shows a strong peak at $72.5\GeV$ owing to the
$\ga\PZ$-production subprocess.    For the CC/NC process $\Pep \Pem \to
\Pu\,\Pubar\,\Pd\,\Pdbar\,\ga$, the electroweak diagrams dominate
below the WW threshold, whereas the gluon-exchange diagrams dominate
at the $\ga\PZ$ peak and above.  The interference between purely
electroweak and gluon-exchange diagrams is generally small.

\section{Summary and outlook}
\label{se:sum}

The class of processes $\Pep\Pem\to 4\,\mbox{fermions} + \ga$ has been
discussed in detail. After classifying the different final states
according to their production mechanism, the sets of all Feynman
graphs are reduced to two generic subsets that are related to the two
basic graphs of the non-radiative processes $\Pep\Pem\to
4\,\mbox{fermions}$.  In this way, all helicity matrix elements are
expressed in terms of two generic functions. Using the
Weyl--van~der~Waerden spinor formalism, we have given compact
expressions for these functions.  We wrote two independent Monte Carlo
programs, both using the multi-channel integration technique and an
adaptive weight optimization procedure to reduce the Monte Carlo
error.  The results obtained with the two Monte Carlo programs agree
well within the integration error.

The detailed discussion of numerical results comprises a survey of
integrated cross sections and photon-energy spectra for all different
final states. Moreover, we have numerically compared different ways to
introduce finite decay widths of the massive gauge bosons. Similar to
the known results for $\Pep\Pem\to 4\,\mbox{fermions}$, we find that
the application of running gauge-boson widths leads to totally wrong
results for the radiative processes. Using constant widths
consistently, leads to meaningful predictions.  For the considered
observables, the results for constant widths practically coincide with
those obtained in a complex-mass scheme that fully respects gauge
invariance.  In the latter scheme gauge-boson masses are treated as
complex parameters everywhere, in particular, leading to complex
couplings.  Similar to the situation for the well-known non-radiative
processes, we find that for precise predictions the diagrams
corresponding to triple-gauge-boson-production subprocesses are not
sufficient and the inclusion of the other graphs is mandatory.
Finally, we have investigated the relevance of gluon-exchange
contributions. In general, both the purely electroweak contributions
and the gluon-exchange contributions are relevant and either one can be
dominant depending on the process and the considered observable.  The
interference of both contributions is at most at the per-cent level.

In this paper, we have assumed that the radiated photon appears as a
detectable particle in the final state, i.e.\ soft and collinear photons
are excluded by cuts. The inclusion of soft and collinear photons is,
however, necessary if $\Pep\Pem\to 4\,\mbox{fermions}+\ga$ is considered
as a correction to four-fermion production. The analytical results of
this paper and the constructed Monte Carlo programs 
will be used as a building block in the 
evaluation of four-fermion production in $\Pep\Pem$ collisions 
including ${\cal O}(\alpha)$ corrections. 

\appendix
\section*{Appendix}

\section*{An example for phase-space generation}
\renewcommand{\theequation}{A.\arabic{equation}}

To illustrate the phase-space generation in one channel,
we choose the multi-peripheral diagram, shown in \reffi{fi:example}. We
determine the momentum configuration
$k_i$ from the phase-space variables $s_i$, $t_i$, $\phi_i$, and
$\theta_i$
with a suitable choice of mappings 
and calculate the resulting probability density $g$. 
\begin{figure}
\begin{center}
\begin{picture}(150,110)(0,0)
\ArrowLine(50,105)(0,105)
\ArrowLine(0,5)(50,5)
\ArrowLine(150,105)(50,105)
\ArrowLine(50,5)(150,5)
\Photon(50,105)(50,55){2}{4}
\Photon(50,5)(50,55){2}{4}
\Photon(100,55)(50,55){2}{4}
\ArrowLine(100,55)(125,70)
\ArrowLine(125,70)(150,85)
\ArrowLine(150,25)(100,55)
\Photon(125,70)(150,55){-2}{3}
\Vertex(50,105){2}
\Vertex(50,5){2}
\Vertex(50,55){2}
\Vertex(100,55){2}
\Vertex(125,70){2}
\put(-40,100){\makebox(0,0)[lb]{$\Pe^+(p_+)$}}
\put(-40,0){\makebox(0,0)[lb]{$\Pe^-(p_-)$}}
\put(10,75){\makebox(0,0)[lb]{$W(q_2)$}}
\put(10,25){\makebox(0,0)[lb]{$W(q_1)$}}
\put(55,60){\makebox(0,0)[lb]{$Z(k_{345})$}}
\put(155,100){\makebox(0,0)[lb]{$\bar\nu_\Pe(k_1)$}}
\put(155,0){\makebox(0,0)[lb]{$\nu_\Pe(k_2)$}}
\put(155,80){\makebox(0,0)[lb]{$\mu^-(k_3)$}}
\put(155,20){\makebox(0,0)[lb]{$\mu^+(k_4)$}}
\put(155,50){\makebox(0,0)[lb]{$\gamma(k_5)$}}
\end{picture}
\end{center}
\caption{
An example for a multi-peripheral diagram contributing to the
process $\Pep\Pem\to \Pne\Pnebar \mu^- \mu^+ \gamma$}
\label{fi:example}
\end{figure}

First we study the propagator structure of the Feynman diagram 
of \reffi{fi:example} and choose a set $\vec{\Phi}$ 
of 11 phase-space variables that is suitable for importance sampling.
Our choice of $\vec{\Phi}$ consists of 
three time-like invariants 
\begin{eqnarray}
s_{35} &=& k_{35}^2 = (k_3+k_5)^2,
\nonumber \\
s_{345} &=& k_{345}^2 = (k_3+k_4+k_5)^2,
\nonumber \\
s_{1345} &=& k_{1345}^2 = (k_1+k_3+k_4+k_5)^2,
\end{eqnarray}
two space-like invariants
\begin{eqnarray}
t_1 &=& q_1^2 = (p_- - k_2)^2,
\nonumber \\
t_2 &=& q_2^2 = (p_+ - k_1)^2,
\end{eqnarray}
and six
 azimuthal and polar angles $\phi_{1,2,3,4}$ and $\theta_{3,4}$, 
respectively.
As worked out in detail in \citere{By73}, the 11-dimensional 
phase-space integral in \refeq{eq:crosssection} can
be written as follows:
\begin{eqnarray}
R_5(s)&=&\int \left[\prod\limits_{i=1}^5 \rd^4 k_i \, 
\delta\left(k_i^2\right) \theta(k_i^0)\right]
\delta^{(4)} \left( p_+ +p_- -\sum_{i=1}^5 k_i \right)  
\nonumber \\
&=& \int_V \rd s_{35} \, \rd s_{345} \, \rd s_{1345} 
\,R_2(s) \, R_2(s_{1345}) \, R_2(s_{345}) \, R_2(s_{35}),  
\end{eqnarray}
where $\sqrt{s}$ denotes the CM energy with $s=(p_- + p_+)^2$.
The phase space of the $2 \to 5$ scattering process 
is composed of the phase spaces of two $2 \to 2$ scattering processes 
with two subsequent
$1 \to 2$ decays described by the 
following functions:
\begin{description}
\item[$R_2(s)$:]
First, we consider the $2\to 2$ scattering process 
$p_+ + p_- \to k_{1345} + k_2$ in the system $\vec{p}_+ + \vec{p}_- =\vec{0}$
where the direction of $\vec{p}_+$ is chosen to be the positive $z$ axis.
\item[$R_2(s_{1345})$:]
Then, we 
boost and rotate to the system 
$\vec{k}_{1345}^{\, \prime} =\vec{0}$
and describe the $2\to 2$ scattering process 
$p_+^{\prime}+ q'_1 \to k_{345}^{\prime} + k_1^{\prime}$, 
where the direction of $\vec{p'}_+$ 
is chosen to be the positive $z$ axis. 
\item[$R_2(s_{345})$:]
Next, we define the $1\to 2$ decay of the Z~boson
$k_{345}^{\prime \prime} \to k_{35}^{\prime \prime}+ k_4^{\prime \prime}$ 
in its rest frame $\vec{k}_{345}^{\, \prime \prime}=\vec{0}$.
\item[$R_2(s_{35})$:]
Finally, we treat the subsequent $1\to 2$ decay of the virtual muon  
$k_{35}^{\prime \prime \prime}\to 
k_3^{\prime \prime \prime}+k_5^{\prime \prime \prime}$ in its rest frame
$\vec{k}_{35}^{\prime \prime \prime}=\vec{0}$.
\end{description}
For massless external fermions, $R_5(s)$ explicitly reads%
\footnote{As usual we set $f=0$ outside the imposed cuts. However, to
  improve the efficiency of the phase-space generation, cuts can be
  included already in the physical boundaries of phase space. In our
  example we have introduced a lower cut on $s_{35}$.}
\begin{eqnarray}
R_5(s)&=&\int_{s_{35,\mathrm{cut}}}^s \rd s_{35} \int_{s_{35}}^s \rd s_{345} 
\int_{s_{345}}^s \rd s_{1345} 
\left[\frac{1}{4 s} \int_{t_{1,\min}}^{t_{1,\max}} \rd t_1 \int_{0}^{2 \pi} 
\rd \phi_1 \right]
\nonumber \\
&& \times   \left[\frac{1}{4 \lambda^{1/2}(s_{1345},t_1,0)} 
\int_{t_{2,\min}}^{t_{2,\max}} \rd t_2 \int_{0}^{2 \pi} \rd \phi_2 \right]
\nonumber \\
&& \times  \left[\frac{\lambda^{1/2}(s_{345},s_{35},0)}{8 s_{345}} 
\int_{-1}^{1} \rd \cos\theta_3 
\int_{0}^{2 \pi} \rd \phi_3 \right] \, 
\left[\frac{1}{8} \int_{-1}^{1} \rd \cos\theta_4 \int_{0}^{2 \pi} 
\rd \phi_4 \right] 
\end{eqnarray} 
with the kinematical function 
\begin{equation}\label{eq:kinematic}
\lambda(x,y,z)=x^2+y^2+z^2-2xy-2yz-2zx.
\end{equation}
The limits on $t_{1,2}$ are the physical boundaries 
of the $2 \to 2$ scattering processes
\begin{eqnarray}
\begin{array}{rlrl}
t_{1,\min}&=s_{1345}-s, 
& t_{1,\max}&=0,
\nonumber\\
t_{2,\min}&=(t_1-s_{1345})(s_{1345}-s_{345})/s_{1345}, \qquad
& t_{2,\max}&=0.
\end{array}
\end{eqnarray}
In our example, we only have to cope with massive space-like
propagators, so that we generate the invariants $s_{35}$ and $s_{345}$
according to \refeqs{eq:maps1} and \refeqf{eq:maps2}, but do not apply
importance sampling in the variables 
$s_{1345}$, $t_{1,2}$, $\phi_{1,2,3,4}$, and $\cos\theta_{3,4}$, 
which are generated according to \refeq{eq:nomaps}.
The resulting density $g$ of \refeq{eq:mdens} for this specific
channel reads for $\nu\neq 1$,
\begin{eqnarray}
\frac{1}{g} &=& 
\frac{\lambda^{1/2}(s_{345},s_{35},0)}{4^2\, 8^2 \, s \, s_{345}
\, \lambda^{1/2}(s_{1345},t_1,0)}   
\nonumber\\
&& \times  s_{35}^{\nu} \, 
\frac{s^{1-\nu}-s_{35,\mathrm{cut}}^{1-\nu}}{1-\nu}\,
\frac{[(s_{345}-\MZ^2)^2+\MZ^2 \GZ^2]}{\MZ \GZ}\, (y_2-y_1)
\nonumber\\
&&\times  (s-s_{345})\, |t_{1,\min}|\, |t_{2,\min}|\, 4\, (2 \pi)^4 
\end{eqnarray}
with $y_{1,2}=\arctan[(s_{1,2}-\MZ^2)/(\MZ\GZ)]$, 
$s_1=s_{35}$, and $s_2=s$.
Finally, the event characterized by the four-momentum configuration
$k_i, i=1,\ldots ,5,$ 
and the beam momenta are defined in terms of 
the generated invariants and angles as follows:
\begin{eqnarray}
p_+^\mu &=& \frac{\sqrt{s}}{2} (1,0,0,1),
\nonumber \\
p_-^\mu &=& \frac{\sqrt{s}}{2} (1,0,0,-1),
\nonumber \\
k_2^\mu &=& \frac{\lambda^{1/2}(s,s_{1345},0)}{2 \sqrt{s}}
(1,-\cos\phi_1 \sin\theta_1,-\sin\phi_1 \sin\theta_1,-\cos\theta_1)
\end{eqnarray}
with 
$\cos\theta_1=(t_1+2 p_-^0 k_2^0)/(2 p_-^0 k_2^0)$,
\begin{equation}
k_1^{\prime\mu} = \frac{\lambda^{1/2}(s_{1345},s_{345},0)}{2 \sqrt{s_{1345}}}
(1,\cos\phi_2 \sin\theta_2,\sin\phi_2 \sin\theta_2,\cos\theta_2)
\end{equation}
with 
$\cos\theta_2=(t_2+2 p_+^{\prime\,0} k_1^{\prime\,0})/(2 p_+^{\prime\,0} k_1^{\prime\,0})$
and $p_+^{\prime\,0}=(s_{1345}-t_1)/(2\sqrt{s_{1345}})$, 
\begin{equation}
k_4^{\prime \prime\mu}= \frac{\lambda^{1/2}(s_{345},s_{35},0)}{2 \sqrt{s_{345}}}
(1,-\cos\phi_3 \sin\theta_3,-\sin\phi_3 \sin\theta_3,-\cos\theta_3),
\end{equation}
and
\begin{eqnarray}
k_3^{\prime \prime \prime\mu} &=& \frac{\sqrt{s_{35}}}{2}
(1,\cos\phi_4 \sin\theta_4,\sin\phi_4 \sin\theta_4,\cos\theta_4), 
\nonumber \\
k_5^{\prime \prime \prime\mu}&=& (k_3^{\prime \prime \prime\,0}, 
- \vec{k}_3^{\prime \prime \prime}).
\end{eqnarray}
The four-momenta $k_1$, $k_3$, $k_4$, and $k_5$ in the CM system
are obtained from
the four-momenta $k_1^\prime$, $k_3^{\prime \prime \prime}$,
$k_4^{\prime \prime}$, and $k_5^{\prime \prime \prime}$ by applying
the appropriate Lorentz transformations, as for instance
described in \citere{By73}.

\section*{Acknowledgement}
We thank D. Graudenz and R. Pittau for useful discussions about
Monte Carlo integration and T. Stelzer for discussions concerning
Madgraph.

\end{document}